\pdfoutput=1
\documentclass[a4paper,12pt]{article}
\usepackage[utf8]{inputenc}
\usepackage{amssymb,amsmath,amsfonts,bbm,dsfont,graphicx,hyphenat,bm,float,a4wide,xcolor}
\usepackage[english]{babel}
\usepackage[colorlinks,allcolors=blue]{hyperref}
\usepackage[square,numbers,sort&compress]{natbib}
\usepackage[caption=false]{subfig}

\numberwithin{equation}{section}

\newcommand{\beq}{\begin{eqnarray}}
\newcommand{\eeq}{\end{eqnarray}}
\newcommand{\non}{\nonumber\\}

\DeclareMathOperator{\Og}{O}
\DeclareMathOperator{\SO}{SO}

\newcommand{\p}{\partial}
\renewcommand{\i}{\mathrm{i}}
\renewcommand{\d}{\mathop{}\!\mathrm{d}}

\newcommand{\calQ}{\mathcal{Q}}

\newcommand{\bea}{\begin{eqnarray}}
\newcommand{\eea}{\end{eqnarray}}
\newcommand{\be}{\begin{equation}}
\newcommand{\ee}{\end{equation}}

\DeclareMathOperator{\nm}{nm}
\DeclareMathOperator{\Tesla}{T}
\DeclareMathOperator{\Joule}{J}

\begin{document}
\title{
\vskip 20pt
\bf{Magnetic Skyrmions: \\
  from lumps to supercompactons}
}
\vskip 40pt  

\author{
Stefano Bolognesi$^{(1, *)}$, \ Sven Bjarke Gudnason$^{(2, \dagger)}$, \ Roberto Menta$^{(3, \ddagger)}$ 
\\[13pt]
  {\em \footnotesize
    $^{(1)}$Department of Physics ``E. Fermi", University of Pisa, and INFN, Sezione di Pisa}\\[-5pt]
  {\em \footnotesize
    Largo Pontecorvo, 3, Ed. C, 56127 Pisa, Italy}\\[2pt]
  {\em \footnotesize	$^{(2)}$Institute of Contemporary Mathematics, School of Mathematics and Statistics,
  }\\[-5pt]
  {\em \footnotesize
    Henan University, Kaifeng, Henan 475004, P.~R.~China
  }\\[2pt]
  {\em \footnotesize
    $^{(3)}$Scuola Normale Superiore, Piazza dei Cavalieri, 7, and Laboratorio NEST, 
    }\\[-5pt]
  {\em \footnotesize
    Piazza S.~Silvestro, 12, 56127  Pisa, Italy}\\[2pt]
  \\[1pt]
{\footnotesize  $^{(*)}$\texttt{stefano.bolognesi@unipi.it}, \ \  $^{(\dagger)}$\texttt{gudnason@henu.edu.cn},  \ \  $^{(\ddagger)}$\texttt{roberto.menta@sns.it}}
}

\date{}

\vskip 8pt
\maketitle

\begin{abstract}
The magnetic Skyrmion is described by one control parameter and one length
scale. We study the two extreme limits of the control parameter -- infinitely large and vanishing -- and find that the magnetic
Skyrmion becomes a ``restricted'' magnetic Skyrmion and an $\Og(3)$
sigma model lump, respectively.
Depending on the potential under consideration, the restricted limit
manifests differently. In the case of the Zeeman term, the restricted
magnetic Skyrmion becomes a ``supercompacton'' that develops a
discontinuity, whereas for the Zeeman term to the power 3/2 it becomes
a normal compacton.
In both the lump and the restricted limit the solution is given in
exact explicit form.
We observe that the case of the Zeeman term squared, which
can also be understood as a special combination of the Zeeman term and
the easy-plane potential -- realizable in the laboratory,
the analytically exact solution for all values of the coupling --
including the Bogomol'nyi-Prasad-Sommerfield (BPS) case -- is also of the lump type.
Finally, we notice that certain materials (e.g.~Fe$_{1-x}$Co$_x$Si or
Mn$_{1-x}$Fe$_x$Ge) have a rather large control parameter $\epsilon$
of order 100, making the restricted limit a suitable rough
approximation.
\end{abstract}

\newpage
\tableofcontents

\section{Introduction}\label{intro}

Topological soliton applications cover many fields of theoretical
physics, from high energy physics to condensed matter theory
\cite{rajaraman1989introduction, MantonSutcliffe, Braun2012,
  shifman2012advanced, Shnir_2018, han2017skyrmions}.
Some topological solitons that have garnered significant
attention in the field of condensed matter physics in recent years are
known as \emph{magnetic Skyrmions} named after the famous works of
T.~Skyrme \cite{skyrme1961non, skyrme1962unified}.
These magnetic Skyrmions are the main focus of this work.
From a condensed matter point of view, a magnetic Skyrmion is a
stable two-dimensional nanoparticle describing a localized winding of
the magnetization in certain magnetic materials \cite{belavin1975i,
  bogdanov1989thermodynamically, bogdanov1994thermodynamically}.
These topological defects are subject of intense experimental and
theoretical investigations \cite{roessler2006spontaneous,
  Rossler:2010st, Ezawa:2010uy, Banerjee:2014hna, melcher2014chiral,
  Rybakov:2018bxt, schroers2019gauged, barton2020magnetic,
  kuchkin2020magnetic, ross2021skyrmion, hill2021chiral,
  schroers2021solvable, Amari:2022boe, Hanada:2023lnm} and have shown
potential for technological spintronic applications
\cite{fert2013skyrmions, fert2017magnetic}.
They were observed for the first time in 2009
by S.~Muhlbauer \emph{et al}.~\cite{muhlbauer2009skyrmion} in a chiral
magnet of MnSi, and subsequently in a plethora of other chiral
magnets~\cite{Munzer2010, Yu2010, Seki2012, Yu2011,
  nagaosa2013topological, Tokura2021}.
Magnetic Skyrmions have risen to prominence as up-and-coming
candidates for next-generation high-density efficient information
encoding \cite{everschor2018perspective,
  luo2021skyrmion} or for quantum computing applications
\cite{psaroudaki2023skyrmion}. 

The spin texture of a magnetic Skyrmion is a stable configuration that
originates from chiral interactions, known as Dzyaloshinskii–Moriya
(DM) interactions. The Hamiltonian interaction term involving two
atomic spins in the chiral magnet, was proposed by Dzyaloshinskii
\cite{dzyaloshinsky1958thermodynamic} as an additional contribution to
the usual Heisenberg Hamiltonian. Subsequently, Moriya
\cite{moriya1960anisotropic, moriya1960new} demonstrated that this
type of interaction originates from the relativistic spin-orbit
coupling. It was suggested \cite{Bak_1980} to consider the mean field approximation
(MFT) for the field theory model in the continuum limit, where
both the Heisenberg exchange term (also called Dirichlet term, $E_2$)
and the DM one ($E_1$) are competing and the Zeeman energy (the
external potential, $E_0$) is included to control the system.

In this work we consider the theory of magnetic Skyrmions in two
spatial dimensions (2D) and apply certain techniques used in the past
in the context of the baby-Skyrme model \cite{Bolognesi:2014ova}. 
We show that a dimensionless parameter enables us to control the
relation between the Dirichlet term, the one with two derivatives, and
the rest of the energy functional that contains the potential and 
the DM term. By changing this parameter we can flow from an almost
Bogomol'nyi-Prasad-Sommerfield (BPS)
theory, close to the $\Og(3)$ sigma model which admits lumps as
solutions, to a ``restricted'' magnetic Skyrmion model with just $E_1$
and $E_0$. The restricted model is analytically solvable. In the case
of the ordinary Zeeman potential, which is quadratic in the
perturbative excitation fields
near the vacuum, a feature never encountered before in the
context of the baby-Skyrme model is present: the solution is not only a
compacton\footnote{
  Compactons are solitonic solutions with compact support: they have a
  nontrivial profile function for $r<R$ and vanishing energy outside,
  i.e.~for $r>R$. Often the fields are continuous but their
  derivatives are not at $r=R$.
  Compactons exist in the baby-Skyrme model
  \cite{Gisiger:1996vb,Adam:2009px,Adam:2010jr}, the BPS-Skyrme 
  model \cite{Adam:2010fg}, as Q-balls in the
  gauged (ungauged) $\mathbb{C}P^N$ model
  \cite{Klimas:2023zxm}(\cite{Klimas:2017eft}) and in K-field theories
  \cite{Adam:2008rf}, etc. 
}
but it is also discontinuous. We may call it a
``supercompacton''. The discontinuity is resolved as soon as we turn
on a small coefficient for the Dirichlet term.
We also confirm the result of
Ref.~\cite{kuchkin2020magnetic}, that the magnetic Skyrmion being a
solution to the variational equations of motion only corresponds to a
stationary point of the energy functional if a certain boundary term
is subtracted off. This boundary term can be written in a form of a
vorticity \cite{barton2020magnetic,kuchkin2020magnetic}.
Indeed, this proves important for determining the correct lump sizes
in the limit of the control parameter ($\epsilon$) going to zero.

Finally, we observe that this lump-type solution corresponds to the
solution of the critically coupled version of magnetic Skyrmions with
the Zeeman term squared, as described by Barton-Singer~\emph{et
al}.~\cite{barton2020magnetic}.
It is not limited to this scenario: it is also the solution
to the full second-order variational equation of motion for \emph{all}
values of the coupling, first observed by D{\"o}ring and Melcher
\cite{Doring2017}.
We find that for this potential, discarding the boundary term changes
the energy functional to become dependent on $\epsilon$, hence
revealing a phase transition between the homogeneous phase and the
Skyrmionic phase.

On the experimental side, we also confront the dimensionless parameter
with the various experimental realizations of the magnetic Skyrmion
and find that the (near) ``restricted'' magnetic Skyrmion --
corresponding to the case of a large control parameter -- is actually
realized in some materials, like Fe$_{1-x}$Co$_x$Si or Mn$_{1-x}$Fe$_x$Ge.
We also confirm that it is possible to experimentally realize the
Zeeman squared term, by using magnets with a certain anisotropy term
and adjusting the magnetic field of the Zeeman term, so as to complete
the square.
At this critically tuned value of the magnetic field, the exact
analytic lump-type solution is thus realizable in the laboratory.

This paper is organized as follows.
In Sec.~\ref{sec:2D} we review the magnetic Skyrmion in the continuum
formulation with various potentials in our notation.
In Sec.~\ref{sec:hedgehog}, we review the hedgehog Ansatz and discuss
the symmetry breaking properties leading to the fixing of the phase
parameter.
In Secs.~\ref{sec:lump} and \ref{sec:restricted} we introduce the
sigma model lump and restricted limits of the magnetic Skyrmion.
In Sec.~\ref{sec:critical} we review the case of the critical
coupling and in Sec.~\ref{sec:any_coupling} we show that the lump
solution is a solution to the model with the Zeeman term squared for
any coupling.
In Sec.~\ref{sec:numerical} we provide numerical solutions that
interpolate between the lump and restricted limits, verifying the
analytical results obtained.
In Sec.~\ref{sec:exp} we compare the control parameter to experimental
parameters of various chiral magnets.
We conclude in Sec.~\ref{sec:conclusion} with a discussion.

\section{2D Magnetic Skyrmions}\label{sec:2D}

A magnetic Skyrmion is a stable \emph{vortex-like} configuration of
magnetic spin vectors~\cite{belavin1975i,bogdanov1989thermodynamically,bogdanov1994thermodynamically}.
The relevant physical quantity in a 2-dimensional magnet is the
direction of the magnetization vector, $\bm{n}$. The latter is the
order parameter for the system of spins $\bm{S}_i$, localized on the
sites of 2-dimensional lattices. The simplest model describing such
a system is given by the Heisenberg Hamiltonian 
\begin{equation}
  \mathcal{H}_{\mathrm{H}} = -\sum_{i<j} \mathbb{J}_{ij} \ \bm{S}_i \cdot \bm{S}_j ,
\end{equation}
where the interaction is only between nearest neighbor's spins, and
$\mathbb{J}_{ij}$ is the exchange interaction constant (which is
negative in ferromagnetic systems).
We suppose that $\mathbb{J}_{ij}=J$ for $i$ and $j$ being nearest
neighbor sites in the crystal.
In the mean field approximation, a proper dynamical variable is the
expectation value of the spins, i.e.~the unit magnetization vector
$\bm{n}\in S^2$. Its dynamics is governed by the continuum
Ginzburg-Landau (GL) effective theory (see
e.g.~Ref.~\cite{han2017skyrmions}). This is the static energy of the
$\Og(3)$ sigma model. Furthermore, the Hamiltonian of the Heisenberg
model can be supplemented by the potential-like term
$-\bm{B}\cdot\bm{n}$ that describes the Zeeman interaction with an
external magnetic field, which we choose to be aligned along the third
Cartesian axis, $\bm{B}=B\bm{e}_3$.

In cubic crystals, without an inversion center, such as MnSi, symmetry
analysis reveals an additional term that can be incorporated into the
GL description, known as the Dzyaloshinskii-Moriya (DM) interaction~\cite{dzyaloshinsky1958thermodynamic, moriya1960anisotropic,
  moriya1960new}: 
\begin{equation}\label{Hamilt DM}
  \mathcal{H}_{\mathrm{DM}} = -\sum_{ij} \bm{D}_{ij} \cdot (\bm{S}_i \times \bm{S}_j) ,
\end{equation}
where $\bm{D}_{ij}$ is the chiral coupling parameter, chosen
here to be equal to $D\hat{\bm{e}}_{j-i}$ for $i$ and $j$ being
nearest neighbor sites in the crystal, $\hat{\bm{e}}_{j-i}$ is a
unit vector pointing from the lattice site $i$ to the lattice site
$j$ and $D$ depends on the material.\footnote{The chosen DM term here is
denoted the Bloch-type DM term in the literature
\cite{han2017skyrmions,barton2020magnetic,ross2021skyrmion}. }
The corresponding energy in the continuum limit (see
App.~\ref{appendix}), at the leading order in the long-wavelength
approximation, reduces to the functional:
\begin{equation}\label{energy-functional2D-RIGHT}
  \begin{array}{rlc}
    E[\bm{n}] &= \displaystyle \int \d^2x \; J \left( \dfrac{1}{2}\partial_i \bm{n} \cdot \partial_i \bm{n} +\kappa\epsilon_{iab} \partial^i n^a n^b + h(1-n_3) \right)\\
    &\equiv E_2 + E_1 + E_0 ,
  \end{array}
\end{equation}
where $\bm{n}=(n^1,n^2,n^3)$ is a unit magnetization vector
($\bm{n}\cdot\bm{n}=1$),
$i=1,2$ are planar indices, $a,b=1,2,3$ are (target) spatial indices, 
$\kappa = \frac{2D}{J}$ is the effective DM coupling and
$h=\frac{|B|}{J}$ is the parameter playing the role of the mass or
Zeeman coupling strength.
The DM term, $E_1$, breaks parity but is invariant under a
simultaneous $\Og(3)$ rotation of spin and $\Og(2)$ rotation of
spatial coordinates, a property inherited from its origin as a
spin-orbit interaction. 

In presence of a generic potential term $E_0$, the energy functional
takes the form  
\begin{align}
E[\bm{n}] = \int \d^2x \bigg(&\frac{1}{2} \partial_i\bm{n}\cdot\partial_i\bm{n} +\kappa\epsilon_{iab}   \partial^i n^a n^b + V_p(\bm{n}) \non&+ \frac{\lambda}{2}(\bm{n}\cdot\bm{n} - 1) \bigg),\label{energy-funct2D-general}
\end{align}
where we have introduced the Lagrange multiplier, $\lambda$, to
enforce the unit-length constraint ($\bm{n}\cdot\bm{n}=1$) and we
simply measure the energy in units of $J$.
We choose to work with the class of potentials
$E_0 = \int V_p(\bm{n})\d^2x$, that are parametrized by
\beq\label{generic-potential}
V_p(\bm{n})  =   h \left(1 - \bm{n}\cdot\bm{N}\right)^p,
\eeq
which corresponds to the usual massive Zeeman term for $p=1$.
Note that $\bm{N} = N^a\bm{e}_a$ is the vacuum field.
The perturbative fields have mass $m=\sqrt{h}$ if $p=1$ and $m=0$ if $p>1$. 
We will concentrate on the cases $p=1,\frac32,2$ in this paper.
The case $p=2$ corresponds to a particular linear combination of the
Zeeman term and the so-called easy-plane term
\begin{equation}
h \left(1 - \bm{n}\cdot\bm{N}\right)^2 = 
2h(1-\bm{n}\cdot\bm{N}) + h(\bm{n}\cdot\bm{N})^2 - h.
\end{equation}
As was shown in Ref.~\cite{kuchkin2020magnetic}, the energy functional
is only stationary with respect to the solutions to the variational
equations of motion if a certain boundary term is subtracted off:
\begin{align}
  \widetilde{E}[\bm{n}] &= E[\bm{n}] - \Omega[\bm{n}],\label{eq:En_IBP}\\
  \Omega[\bm{n}] &= \kappa\int\d^2x\;\epsilon_{iab}\p^i n^a N^b,
\end{align}
where $\Omega[\bm{n}]$ is called the vorticity.
The vorticity is not part of the DM term, but a similar term that is
locally different, yet giving the same boundary integral for
finite-energy configurations (i.e.~configuration with
$\lim_{|\bm{x}|\to\infty}\bm{n}=\bm{N}$) is
\beq
\widehat\Omega[\bm{n}]=\kappa\int\d^2x\;\epsilon_{ia3}\p^i(n^a n^3).
\label{eq:Omegatilde}
\eeq
Subtracting either the vorticity $\Omega$ or the boundary term
$\widehat\Omega$ from the energy functional $E[\bm{n}]$ will give
rise to a well-defined variational problem. The first option is given
in Eq.~\eqref{eq:En_IBP} as $\widetilde{E}[\bm{n}]$ and we will call the
second option $\widehat{E}[\bm{n}]=E[\bm{n}]-\widehat{\Omega}[\bm{n}]$.

We can write the DM term in an invariant form.
For this we need a tensor $T^{ai}$ such that the DM term can be written as 
\beq
\mathcal{E}_1 = \kappa\, \epsilon_{abc} T^{ai}  \partial_i n^b n^c .
\eeq
The transformations are $\SO(3)$ for the internal indices $a,b,c$ and
$\SO(2)$ for the spatial index $i$. 
Any tensor related by $\SO(3)$ and $\SO(2)$ transformations to this is
essentially equivalent.
Those are the tensors that leave invariant the maximal amount of
symmetry, $\SO(2)_{\rm diag}$.
We can express the tensors of this form in an invariant way. The
choice is essentially determined once we pair $i=1,2$ with two
corresponding orthogonal unit vectors $T^{ai}$ in $\mathbb{R}^3$,
i.e.,
\beq
T^{a1} T^{a1} = T^{a2}T^{a2} =1 , \qquad T^{a1} T^{a2} = 0 .
\label{conditionT}
\eeq
So the meaning of $\bm{T}$ is a pair of orthogonal unit vectors in $\mathbb{R}^3$.
The energy functional \eqref{energy-functional2D-RIGHT} corresponds to the choice
\beq
\bm{T} = 
\begin{pmatrix}
	1&0\\0&1 \\ 0& 0
\end{pmatrix}, \qquad
\bm{N} = 
\begin{pmatrix}
	0\\0 \\ 1
\end{pmatrix},
\eeq
which in indices is $T^{ai}=\delta^{ai}$ and corresponds to the
Bloch-type DM term.
A more general DM term can thus be written using a different tensor
obeying the conditions \eqref{conditionT}. In particular,
Refs.~\cite{barton2020magnetic,ross2021skyrmion} use
\beq
\bm{T} =
\begin{pmatrix}
  \cos\alpha & -\sin\alpha\\
  \sin\alpha & \cos\alpha\\
  0& 0
\end{pmatrix},
\eeq
where $\alpha=0$ corresponds to the Bloch-type DM term and
$\alpha=\frac\pi2$ to the N\'eel-type DM term.

The equations of motion are independent of boundary terms and read
\begin{equation}
\partial_i^2n^a +  h p  N^a  \left(1-\bm{n}\cdot\bm{N} \right)^{p-1}
- 2\kappa\epsilon_{aib}  \partial^i n^b
- \lambda n^a = 0 ,
\end{equation}
$\bm{n}\cdot\bm{n}=1$, where the Lagrange multiplier is
\begin{equation}
\lambda = n^a\left(\partial_i^2n^a
+ h p N^a\left(1- \bm{n}\cdot\bm{N}\right)^{p-1}
- 2 \kappa \epsilon_{aib} \partial^i n^b
\right).
\end{equation}
The vector $\bm{N} = \bm{e}_3 = (0,0,1)$ corresponds to the
vacuum. Due to the following condition
\begin{equation}\label{finite-energy-condition3D}
  \lim_{ r \to\infty} \bm{n}= \bm{N} ,
\end{equation} 
the base space of the vector field $\bm{n}: \mathbb{R}^2 \to S^2$,
can be compactified to the sphere $S^2$.
Therefore, due to the nontrivial homotopy group
$\pi_2(S^2)=\mathbb{Z}$, there are topologically nontrivial solutions
(with finite energy) that can be classified according to the degree of
the map $\bm{n}: S^2 \to S^2$, i.e.~the winding number $Q\in\mathbb{Z}$.
The topological degree or the Skyrmion number, $Q$, is given by
\beq
Q = \int\d^2x\;\calQ
= \frac{1}{4\pi}\int\d^2x\;
\bm{n}\cdot\p_1\bm{n}\times\p_2\bm{n}.
\label{eq:Q}
\eeq
The target space $S^2$ can be parameterized by the angles
$(\vartheta,\varphi)$: 
\begin{equation}\label{3sphericalcoord}
	\bm{n}=\begin{pmatrix}
		\sin\vartheta \cos\varphi \\
		\sin\vartheta \sin\varphi \\
		\cos\vartheta \\
	\end{pmatrix},
	\qquad  \ \  \vartheta \in [0,\pi] , \ \   \ \ \varphi \in [0,2\pi) .
\end{equation}
and the energy functional \eqref{eq:En_IBP} becomes
\begin{align}
\widetilde{E}[\vartheta, \varphi] &= \int \d^2x \, \bigg(
\frac{1}{2}(\partial_i\vartheta)^2
+ \frac12\sin^2\vartheta (\partial_i \varphi)^2
+  h( 1- \cos{\vartheta})^p \non
&
+\kappa(1-\cos\vartheta)\Big[
  \sin\varphi(\p_1\vartheta -\sin\vartheta\p_2\varphi)\non
  &\phantom{+\kappa(1-\cos\vartheta)\Big[\ }
    -\cos\varphi(\p_2\vartheta +\sin\vartheta\p_1\varphi)
    \Big]\bigg),
\label{eq:E_thetaphi_cartesian}
\end{align}
whereas for reference, the energy functional with the boundary term
\eqref{eq:Omegatilde} subtracted off instead simplifies as
\begin{align}
  \widehat{E}[\vartheta, \varphi]
  &= \int \d^2x \, \bigg(
\frac{1}{2}(\partial_i\vartheta)^2
+ \frac12\sin^2\vartheta (\partial_i \varphi)^2
+  h( 1- \cos{\vartheta})^p \non
&\quad
+ \kappa  \sin^2\vartheta \,   \left(
\sin\varphi \,	\partial_1 \vartheta - \cos\varphi\,  \partial_2
\vartheta \right)\bigg).
\end{align}
The simplification occurs because the boundary term
\eqref{eq:Omegatilde} is already contained in the original DM term of
Eq.~\eqref{energy-funct2D-general}.

Derrick's argument allows for the existence of these 2D magnetic
solitons. The idea is to make a rescaling of the field
$\bm{n}(x)\mapsto \bm{n}(x/R)$, and see how the energy terms scale
with $R$, i.e.,
\begin{equation}\label{Derrick-rescaling3D}
  E_{\bm{n}}(R) \equiv E[\bm{n}(x/R)] =  E_{2,\bm{n}}(1)  +  E_{1,\bm{n}}(1) R +  E_{0,\bm{n}}(1) R^2 ,
\end{equation}
where the three terms correspond to the kinetic energy, the DM term
and the potential term, respectively.
If the DM term is negative, $E_{\bm{n}}(R)$ starts, for small $R$,
decreasing linearly with $R$ until it increases quadratically due to
the potential contribution. This change allows for the existence of
stationary points corresponding to 
\begin{equation}
  R_{*} = \dfrac{ E_{1,\bm{n}}(1)}{2  E_{0,\bm{n}}(1)},
  \label{Rstar}
\end{equation}
which is roughly the length scale at which a soliton can be
stabilized.
By using the order of magnitude estimate
\be
E_{2,\bm{n}}(1) \propto 1, \qquad
E_{1,\bm{n}}(1)\propto \kappa, \qquad
E_{0,\bm{n}}(1) \propto h,
\ee
we have
\begin{equation}
  R_* \sim \frac{\kappa}{h}  .
\end{equation}
The length scale of the soliton thus increases with $\kappa$ and
decreases with $h$.
We thus define two new parameters:
\beq\label{sizes}
\ell = \frac{\kappa}{h}  , \qquad
\epsilon = \frac{\kappa^2}{h} ,
\eeq
where $\ell$ controls the size of the Skyrmion \eqref{Rstar} while
$\epsilon$, the dimensionless parameter of the theory, controls the
relation between $E_2$ and $E_1 + E_0$.
We call the latter the control parameter of the theory (i.e.~$\epsilon$).
In order to see the scaling behavior of the theory, we can write
\beq
E_{\bm{n}} = \widetilde{E}_{2,\bm{n}} + \kappa\widetilde{E}_{1,\bm{n}}
+ h\widetilde{E}_{0,\bm{n}},
\eeq
and perform the scaling $\bm{n}(x)\mapsto\bm{n}(x/\ell)$, which
yields
\beq
E_{\bm{n}} = \widetilde{E}_{2,\bm{n}} + \epsilon\widetilde{E}_{1,\bm{n}}
+ \epsilon\widetilde{E}_{0,\bm{n}}.
\label{derrick1}
\eeq
Clearly, in the limit of small $\epsilon$, there is only the kinetic
term, whereas in the limit of large $\epsilon$, there is only a small
contribution from the kinetic term.
In these rescaled coordinates, the soliton size is of order one.
On the other hand, if we instead perform the following rescaling
$\bm{n}(x)\mapsto\bm{n}(x\kappa)$, we obtain
\beq
E_{\bm{n}} = \widetilde{E}_{2,\bm{n}} + \widetilde{E}_{1,\bm{n}}
+ \frac{1}{\epsilon}\widetilde{E}_{0,\bm{n}}.
\label{derrick2}
\eeq
In these rescaled coordinates, it is clear that the model only depends
on $\epsilon$, but now the size of the soliton is no longer of order
one, but of order $\epsilon$.

The symmetry group  $\SO(3)_{\rm int}$ (internal) of the sigma model
is broken to $\SO(2)_{\rm int}$ by the potential, viz.~by the
vector $\bm{N}$ in $E_0$. So, by considering $E_2 + E_0$, we have a
symmetry group of spatial and internal rotations,
$\SO(2)_{\rm spt}\times \SO(2)_{\rm int}$.
The DM term locks together spatial and internal rotations, so only a
diagonal combination of the two remains invariant,
$\SO(2)_{\rm diag}$. Rotations of $\SO(2)_{\rm diag}$ act as: 
\begin{equation}
  \bm{n}(x_1,x_2) \mapsto R_{12}(\zeta)\bm{n}(\cos\zeta x_1 - \sin\zeta x_2, \sin\zeta x_1 + \cos\zeta x_2) ,
\end{equation}
$\zeta \in [0,2\pi)$.
Moreover, there are the two internal parities as a discrete group
$P_{\rm spt} \times P_{\rm int}$. The DM interaction is responsible
for the breaking to the diagonal group $P_{\rm diag}$.
Up to an additional internal rotation of $\pi$, the parity transformations act as:
\begin{equation}
  \bm{n}(x_1,x_2) \mapsto \bar{\bm{n}}(x_1,-x_2) ,\qquad
  \bar{\bm{n}} = (n^1,-n^2,n^3) .
\end{equation} 
Therefore, the total invariance group of the  model is given by
\begin{equation}\label{group2D}
  G = \Og(2)_{\rm diag}  \ltimes T_2 ,
\end{equation}
where $T_2$ corresponds to the 2-dimensional translations,
$\Og(2)_{\rm diag}$ contains the diagonal parity and $\ltimes$ denotes
a semidirect product.

\subsection{Hedgehog solution}\label{sec:hedgehog}

The 2-dimensional base space can be parameterized by the polar
coordinates $(r,\phi)$ where $r>0$ and $\phi \in [0,2\pi]$ as
$x_1+\i x_2=r e^{\i\phi}$.
We seek solutions within the class of hedgehog maps and can express 
the hedgehog Ansatz as
\begin{equation}\label{HedgeHog-ansatz2D}
    \vartheta = \vartheta(r),\qquad
    \varphi = -Q \phi + \delta ,
\end{equation}
where $\vartheta(r)$ is the profile function of the soliton.
Since the field must approach the vacuum $\bm{N}=(0,0,1)$ at spatial
infinity, it satisfies the boundary conditions: $\vartheta(0)=\pi$ and
$\vartheta(\infty)=0$.
The phase $\delta$ corresponds to the internal orientation of the
Skyrmion, i.e.~a generic rotation in $x_1$-$x_2$ plane.
Substituting the Ansatz \eqref{HedgeHog-ansatz2D} into the energy
functional \eqref{energy-functional2D-RIGHT} and integrating over
$\varphi$, we find that
\begin{align}\label{E2dSkyrmion-ansatz}
  E &= 2\pi \int r \d r\; \bigg\{\frac{1}{2}(\vartheta')^2
  + \dfrac{Q^2}{2r^2} \sin^2 \vartheta + h(1 - \cos\vartheta)^p \non
  &
  +\frac{\kappa}{\pi(Q+1)}\sin(Q \pi)\sin(\delta - Q\pi)
  \left(\vartheta' - \dfrac{Q}{2r} \sin(2\vartheta)\right)\bigg\} ,
\end{align}
and the topological charge \eqref{eq:Q} reduces to
\begin{align}
Q &= -\frac{Q}{2}\int\sin(\vartheta)\vartheta'\d r
= -\frac{Q}{2}\int_\pi^0\sin(\vartheta)\d\vartheta\non
&= -\frac{Q}{2}[\cos\vartheta]_0^\pi = Q.
\end{align}
We note that the contribution of the DM interaction energy is
nontrivial only for the rotationally invariant configuration with
topological charge $Q=-1$. Indeed we have that 
\begin{equation}
  \frac{\kappa}{\pi(Q+1)} \sin(Q \pi)\sin(\delta - Q\pi) =
  \begin{cases}
    \kappa\sin\delta & \text{if}\quad Q=-1 , \\ 
    0 & \text{if}\quad Q\neq -1 .
  \end{cases}
\end{equation}
Assuming that the chiral coupling constant $\kappa$ is positive, the
energy functional \eqref{E2dSkyrmion-ansatz} has a global minimum for
the phase $\delta=\pi/2$.
It means that the magnetic Skyrmion possesses an intrinsic internal
orientation. 
The energy functional \eqref{E2dSkyrmion-ansatz} for the profile
function $\vartheta(r)$ in the topological sector $Q=-1$, with the
minimal-energy orientation $\delta = \pi/2$, reduces to 
\begin{align}
  E = 2\pi \int r \d r \bigg(&\dfrac{1}{2}(\vartheta')^2 + \dfrac{\sin^2 \vartheta}{2r^2}  + \kappa \left(\vartheta' + \dfrac{\sin(2\vartheta)}{2r}\right)\non &+ h(1 - \cos\vartheta)^p \bigg),\label{energy-integral2D}
\end{align}
\begin{figure}[!htp]
  \centering
  \includegraphics[width=0.6\linewidth]{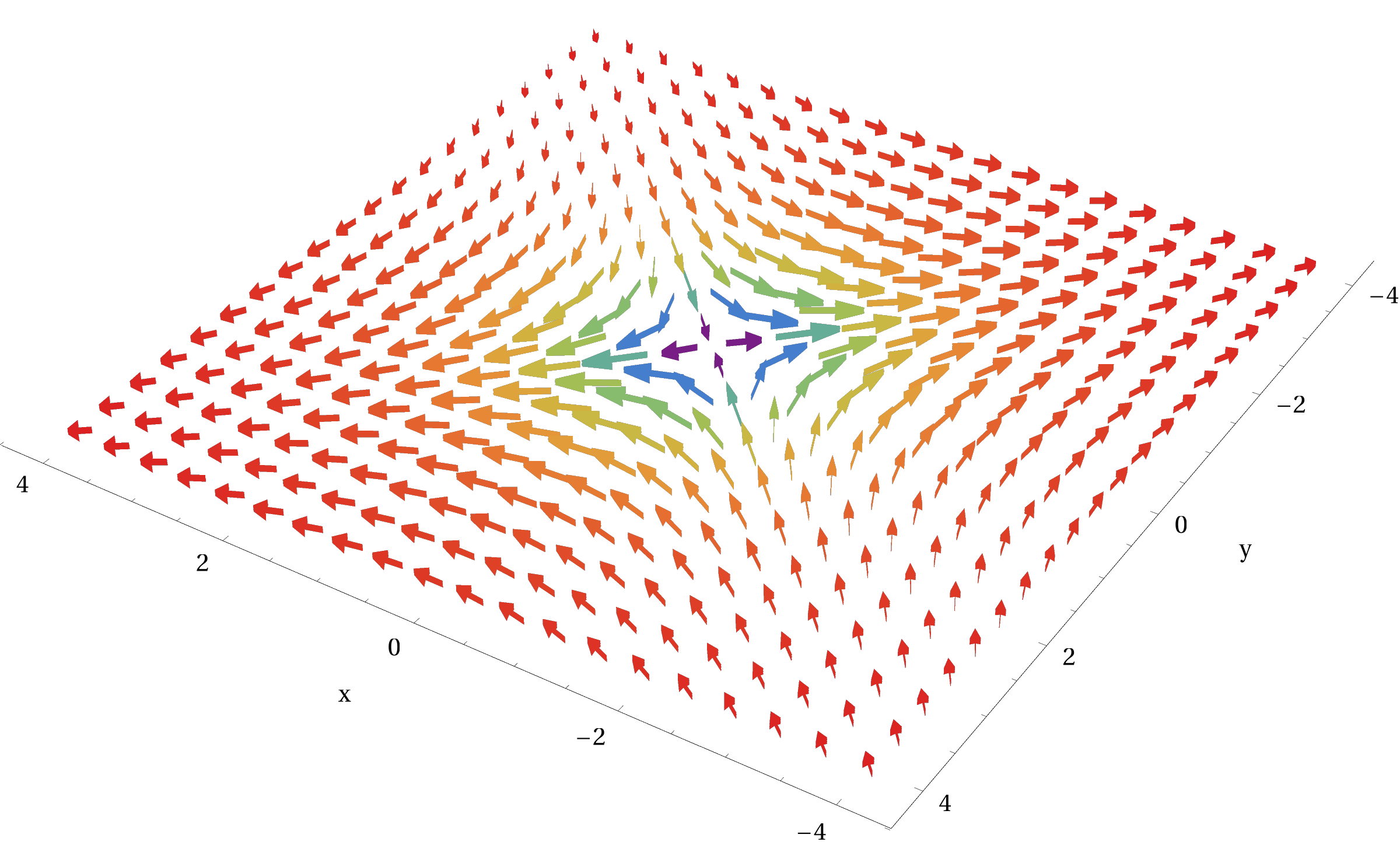}
  \caption{The magnetization vector \eqref{3sphericalcoord} for a typical
      magnetic Skyrmion solution.
  }
  \label{fig:vector}
\end{figure}
and the vector plot of the corresponding magnetization vector
\eqref{3sphericalcoord} is shown in Fig.~\ref{fig:vector}.
The energy functional with the vorticity and boundary term subtracted off
reads
\begin{align}
  \widetilde{E} &= 2\pi \int r \d r \bigg(\frac{1}{2}(\vartheta')^2 + \dfrac{\sin^2 \vartheta}{2r^2} \non& + \kappa (1-\cos\vartheta)\left(\vartheta' - \dfrac{\sin\vartheta}{r}\right)+ h(1 - \cos\vartheta)^p \bigg),\label{eq:Etilde_hedgehog}
  \\
  \widehat{E} &= 2\pi \int r \d r \bigg(\frac{1}{2}(\vartheta')^2 + \dfrac{\sin^2 \vartheta}{2r^2}  + \kappa \sin^2(\vartheta)\vartheta'\non &+ h(1 - \cos\vartheta)^p \bigg),
\end{align}
respectively.
The associated variational equation is
\begin{align}
	\vartheta'' + \dfrac{1}{r}\vartheta' + \frac{2\kappa}{r}\sin^2\vartheta - \dfrac{1}{2r^2} \sin(2\vartheta)\qquad\non - h p \sin\vartheta (1 - \cos\vartheta)^{p-1}= 0 .\label{profile equation 2D}
\end{align}
Remarkably, for the possible existence of these solutions, a negative
contribution to the energy of the DM interaction is a necessity.
Many analytical works studied mathematical properties proving the
theoretical existence of these stable solutions in chiral magnets
\cite{melcher2014chiral, schroers2019gauged, kuchkin2020magnetic,
  barton2020magnetic, walton2020geometry, schroers2021solvable,
  ross2021skyrmion}.
If we had chosen the boundary conditions for $\vartheta$ corresponding
to those of an anti-Skyrmion: $\vartheta(0)=-\pi$ and
$\vartheta(\infty)=0$, the kinetic energy and potential energy would
remain invariant, but the DM contribution to the energy would flip
sign, making the anti-Skyrmion unstable.
We thus understand that the chirality of the DM term selects either
the Skyrmion or the anti-Skyrmion according to the sign of
$\kappa\sin\delta$.

It will prove useful to consider the asymptotic behavior of the field
$\vartheta$ found by linearizing \eqref{profile equation 2D}.
In particular, we find
\begin{align}
  \begin{cases}
    \vartheta = C K_1(\sqrt{h}r), & p=1,\\
    \vartheta = \frac{C}{r}, & p>1,
  \end{cases}
  \label{eq:asymptotic_behavior}
\end{align}
where $C$ is an undetermined constant and $K_1$ is the modified Bessel
function of the second kind, that tends to zero exponentially.
Notice that for the cases with $p>1$, the profile function $\vartheta$
tends to its vacuum value with a power law and hence much slower than
the cases with a Zeeman term.

\begin{figure}[!htp]
  \centering
  \includegraphics[width=0.5\linewidth]{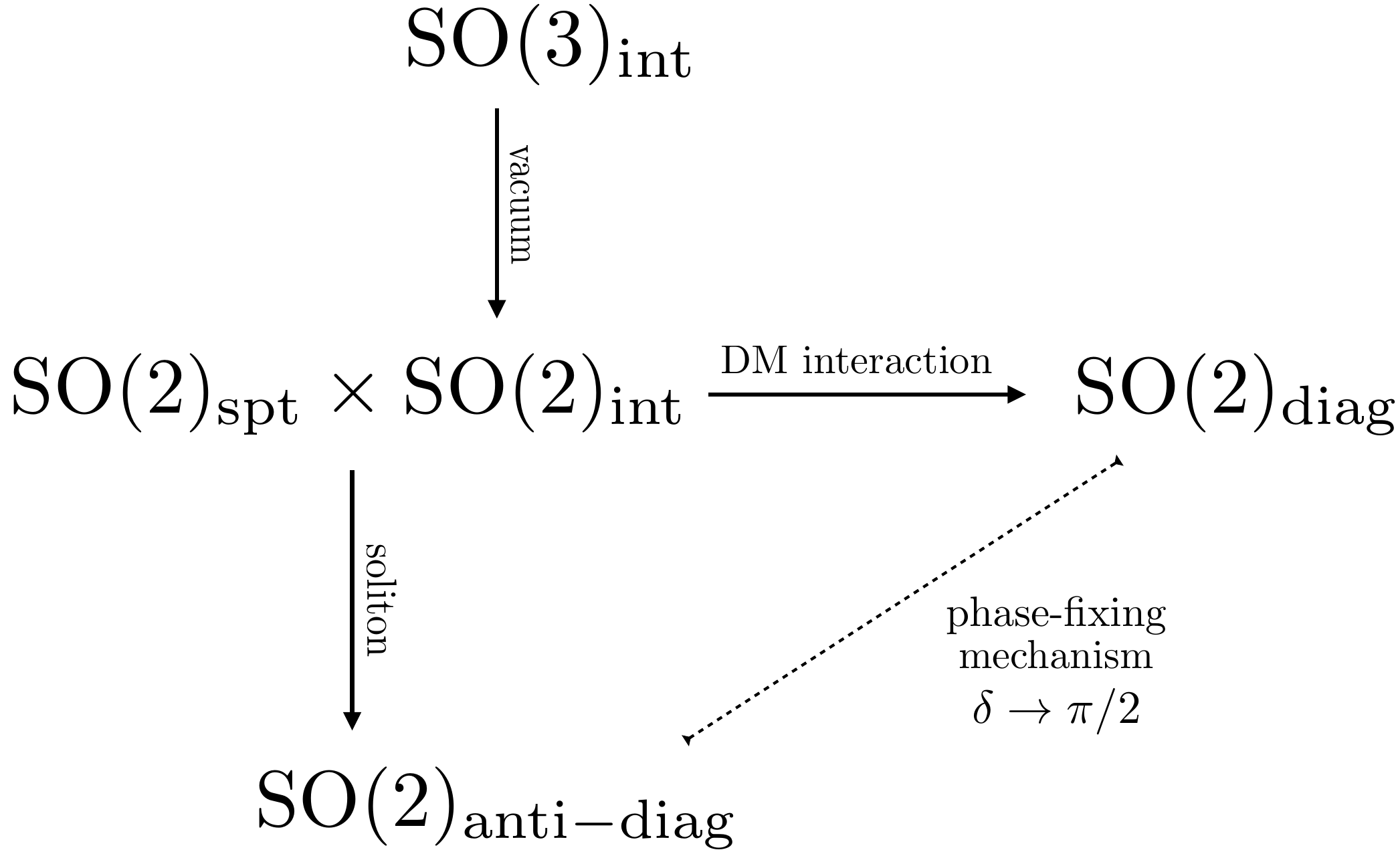}
  \caption{Schematic of the symmetry breaking mechanism for the compact
    symmetry group of the theory.} 
  \label{fig:schemeBS2d}
\end{figure}

At this point, we can analyze the symmetry group of the theory in the
background of the soliton.
The group of translations in the plane $T_2$ of
Eq.~\eqref{energy-integral2D} is broken by the soliton
\eqref{HedgeHog-ansatz2D} and this symmetry breaking is responsible
for two moduli defining the position of the magnetic Skyrmion.
On the other hand, the rotational contribution is more complicated.
If we consider the Lagrangian density in the vacuum without the DM
term, the solitonic background, as we can verify looking at
Eq.~\eqref{HedgeHog-ansatz2D}, breaks the
$\SO(2)_{\rm spt}\times\SO(2)_{\rm int}$ group to
$\SO(2)_{\rm anti-diag}$ (anti-diagonal).
It means that the unbroken group is $\SO(2)$ whose internal angle
would be responsible for another modulus, i.e.~the orientation
$\delta$ of the Skyrmion.
However, the soliton exhibits a similar symmetry-breaking effect as
well in the DM interaction.
Therefore, by considering both the background of the soliton and the
DM interaction together, we observe that further symmetry breaking
occurs for the two separate $\SO(2)_{\rm diag}$ groups, resulting in
the fixing of the orientation angle $\delta$.
As we have previously discussed, the energy is minimized when
$\delta=\pi/2$.
Fig.~\ref{fig:schemeBS2d} shows a scheme of the symmetry-breaking
effect responsible of fixing the $\delta$-phase.
Finally, the moduli space $\mathcal{M}_1$ of a 2-dimensional magnetic
Skyrmion has real dimension
\begin{equation}
  \dim_{\mathbb{R}}(\mathcal{M}_1) = 2 ,
\end{equation}
due only to the translational zeromodes.

\subsection{\texorpdfstring{$\Og(3)$}{O(3)} sigma model limit}\label{sec:lump}

If we keep $\ell \equiv \kappa/h$ fixed and send $\epsilon \equiv
\kappa^2/h \to 0$ we obtain the equation for the sigma model lump with
just the $E_2$ contribution, which has the equation of motion
\begin{equation}
	\vartheta'' + \dfrac{1}{r}\vartheta'   - \dfrac{1}{2r^2} \sin(2\vartheta) = 0 .
\end{equation}
The solution is 
\begin{equation}\label{profilelump}
	\vartheta(r) = 2 \arctan{\frac{a}{r}} ,
\end{equation}
with $a$ an arbitrary scale of the soliton (size modulus of the 
$\Og(3)$ sigma model lump).
We then evaluate the energy functional for small but finite values of
$\epsilon$.
Since it was shown in Ref.~\cite{kuchkin2020magnetic} that $\widetilde{E}$
of Eq.~\eqref{eq:En_IBP} is the energy functional leading to a
well-defined variational problem, we will plug in the above lump
solution to that energy functional:
\beq
\widetilde{E}(a) = 4\pi - 8\pi\kappa a + \frac{2^p}{p-1} h\pi a^2,
\label{eq:Ea}
\eeq
and in turn the lump size
\beq
a = \frac{p-1}{2^{p-2}}\ell.
\label{eq:lumpsize_a}
\eeq
Inserting the lump size back into the energy \eqref{eq:Ea}, we find
\beq
\widetilde{E} = 4 \pi - 4\pi\frac{(p-1)}{2^{p-2}} \epsilon .
\label{eq:Ea_sol}
\eeq
We notice that the correction to the energy is linearly proportional
to $\epsilon$.
This estimate, for $p=2$, predicts that the Skyrmion phase is the
stable phase (compared with the homogeneous ground state) when
$\epsilon>1$.\footnote{
We thank Bruno Barton-Singer for pointing this out to us.
}
For the $p=\frac32$ case, the corresponding critical coupling is
$\epsilon^{\rm crit}=\sqrt{2}$ in the lump approximation (in this
case, the solution is not of lump type for $\epsilon$ large).

For $p=1$ the solution for the lump
\eqref{profilelump} evaluated on $E_0$ diverges, see
Eq.~\eqref{eq:Ea}.
Introducing a cutoff
($\int_0^\infty \d r\longrightarrow\int_0^\Lambda\d r$)
gives 
\beq
\widetilde{E}(a) \simeq 4 \pi - 8 \pi \kappa a +  2 \pi a^2 h \log\left(1+\frac{\Lambda^2}{a^2}\right).
\label{eq:Ea_cutoff}
\eeq
Assuming that $\Lambda\gg a$, we can approximate
$\log(1+\Lambda^2/a^2) \simeq 2\log(\Lambda/a)$.
Hence extremization with respect to $a$ yields
\beq
a = -\frac{\ell}{\omega\left(-\frac{\sqrt{e}\ell}{\Lambda}\right)},
\eeq
where $\omega$ is the Lambert-W function, i.e.~the inverse of the
function $f(W)=We^W$, and $\sqrt{e}\simeq1.649$.
Expanding $a$ in small $\ell=\kappa/h$ yields
\beq
a \simeq \frac{\Lambda}{\sqrt{e}} - \ell + \mathcal{O}(\ell^2\Lambda^{-1}).
\eeq
Unfortunately, it is not possible to remove the infinity leaving
behind a positive results, so this estimate should be considered
unphysical.

Alternatively, we can consider extracting the ``effective'' lump size
from the $\epsilon\to0$ limit, numerically, and regularizing the
energy yielding
\beq
\widetilde{E}_{\rm reg}(a)\simeq 4\pi - 4\pi c \epsilon,
\label{eq:Ep=1reg}
\eeq
with $a=c\ell$.
The above considerations suggest that the lump size $a$ goes to zero
in the $\epsilon\to0$ limit, probably logarithmically or with a slow
power law.

\subsection{Restricted model limit}\label{sec:restricted}

In the case of $\ell$ fixed and $\epsilon \to \infty$ we expect that
the stable Skyrmion is essentially given by a stabilization between
$E_1$ and $E_0$, while $E_2$ is negligible with respect to the other
two terms. Thus the energy functional takes the simplified form\footnote{
One may wonder why we restrict the potential to the form $V_p$ of
Eq.~\eqref{generic-potential} and hence do not consider
e.g.~$V\propto(1-(n^3)^2)$.
Although such easy-axis potential is often utilized in magnetic
systems, they do not allow for the large-$\epsilon$ limit because the
ground state becomes inhomogeneous 
\cite{bogdanov1994thermodynamically,10.21468/SciPostPhys.8.6.086}
and requires a different analysis than carried out here.
}
\begin{equation}
E[\bm{n}]
= \int \d^2 x \left(   \kappa\epsilon_{iab}   \partial^i n^a n^b
+ V_p(\bm{n}) + \frac{\lambda}{2} (\bm{n}\cdot\bm{n} - 1) \right) .
\end{equation}
We may call this model the ``restricted'' magnetic Skyrme model.
The equations of motion are
\beq
hp N^a(1 - \bm{n}\cdot\bm{N})^{p-1}
- 2 \kappa\epsilon_{aib}  \partial^i n^b
- \lambda n^a = 0 ,
\eeq
where the Lagrange multiplier is
\beq
\lambda = n^a\left(  hp N^a(1-\bm{n}\cdot\bm{N})^{p-1} - 2 \kappa\epsilon_{aib} \partial_i n_b
\right) .
\eeq
One interesting aspect of these equations is that they are first order
in the derivative. Thus, they can easily be solved analytically.
Unlike the restricted baby-Skyrme model
\cite{Gisiger:1996vb,Adam:2009px,Adam:2010jr,Bolognesi:2014ova}, here we do not
have an infinite dimensional symmetry group.
The restricted equation for the hedgehog Ansatz has no derivatives and
can be written as 
\begin{equation}
  \dfrac{\kappa}{r}\sin^2\vartheta - \frac{h p}{2} \sin\vartheta (1 - \cos\vartheta)^{p-1}= 0 .
\end{equation}
Solving for $r$ as a function of $\vartheta$, we obtain
\begin{equation}
r = \frac{2\ell\sin\vartheta}{p (1-\cos\vartheta)^{p-1}} ,
\end{equation}
where $\ell=\frac{\kappa}{h}$.
Some examples are:
\begin{align}
  p&=1\, : \quad
  \vartheta(r) =\pi - \arcsin\rho(r),&\quad
  \rho(r) &= \frac{r}{2\ell}, \label{restrp1}\\
  p&=\frac{3}{2}: \quad
  \vartheta(r) = 2\arccos\rho(r), &\quad
  \rho(r) &= \frac{3\sqrt{2}r}{8\ell},\label{restrp3/2}\\
  p&=2\, : \quad
  \vartheta(r) = 2\arctan\frac{1}{\rho(r)},&\quad
  \rho(r) &= \frac{r}{\ell},\label{restrp2}
\end{align}
with $\rho=\rho(r)\in[0,1)$.
\begin{figure}[!htp]
  \centering
  \includegraphics[width=0.5\linewidth]{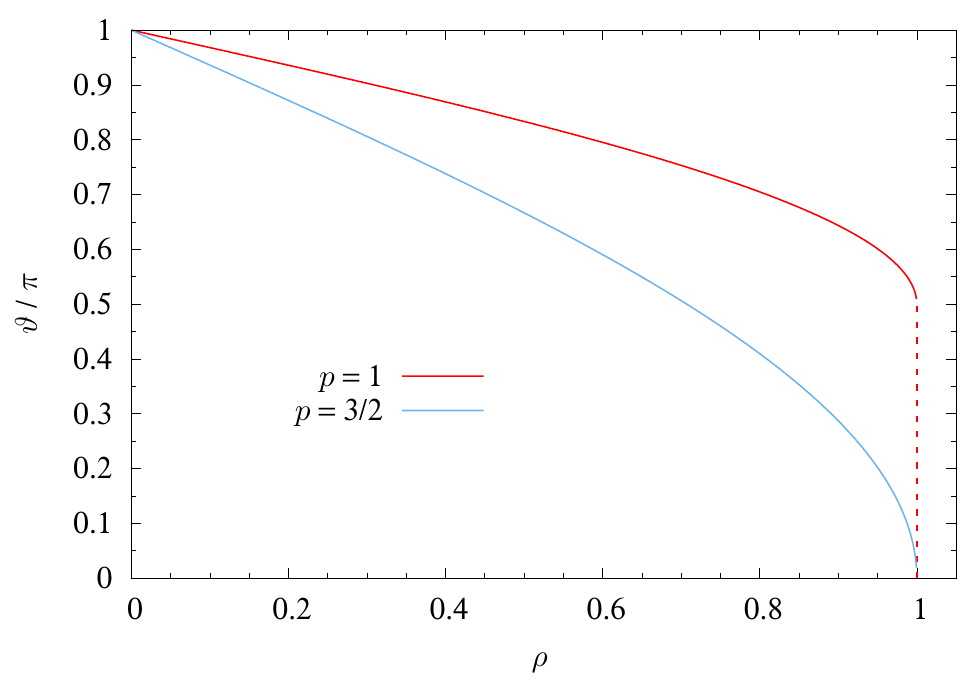}
  \caption{Compacton (blue) for $p=3/2$ \eqref{restrp3/2} 
  \emph{supercompacton} (red) for $p=1$ $\eqref{restrp1}$.
  Notice that $\rho(r)$ has a different numerical prefactor of
  $r/\ell$ in each case. } 
  \label{restricted}
\end{figure}
For $p=1$, an interesting new feature is present in the theory, a
compact and discontinuous solution that we may call a
\emph{supercompacton}.
As shown in Fig.~\ref{restricted}, it differs from the usual compacton
of the baby-Skyrme model by possessing a discontinuity. 
For $p<\frac{3}{2}$, the solution is discontinuous and covers only half of the sphere.
For $p=\frac{3}{2}$ it is a compacton.
For $p>\frac{3}{2}$, $\vartheta(r)$ goes to $0$ smoothly as
$r\to\infty$.

For a generic potential the condition for having a supercompacton is
not necessarily related to the behavior close to the minimum of the
potential. Let us consider, as potential, a generic function
$V(\vartheta)$ with minimum at $\vartheta=0$ whose expansion near the
vacuum is given by $V(\vartheta) = \vartheta^{2p} + \dots$.
The solution of the restricted model, in implicit form, is
\beq
r(\vartheta) = \frac{2\kappa\sin^2\vartheta}{V'(\vartheta)} .
\eeq
A supercompacton is defined as having a discontinuity in the profile function
$\vartheta(r)$. This condition is met when the function $r(\vartheta)$
reaches a maximum at $\vartheta>0$. Consequently, it is not
necessarily related to the power-law near the vacuum. In the event
that the function $r(\vartheta)$ is monotonic and attains a
(finite) maximum at $\vartheta=0$, the restricted solution is not a
supercompacton.

\subsection{Critical coupling}\label{sec:critical}

The critical coupling for the case of $p=2$ can readily be seen from a
Bogomol'nyi completion \cite{barton2020magnetic}
\begin{align}
  E &= 2\pi \int r \d r \left[\dfrac{1}{2}(\vartheta')^2 + \dfrac{\sin^2 \vartheta}{2r^2}  + \kappa \left(\vartheta' + \dfrac{\sin(2\vartheta)}{2r}\right) + h(1 - \cos\vartheta)^2 \right]\non
&= 2\pi\int r\d r\left[
  \frac12\left(\vartheta'
  -\frac{\sin\vartheta}{r}
  +\kappa(1-\cos\vartheta)\right)^2
  +\frac{\sin\vartheta}{r}\vartheta'
  +\frac{\kappa}{r}\p_r(r\sin\vartheta)
  \right]\non
&=\pi\int r\d r
  \left(\vartheta'
  -\frac{\sin\vartheta}{r}
  +\kappa(1-\cos\vartheta)\right)^2
  +4\pi Q + 4\pi\Omega,
  \label{eq:E_Bogomolnyi}
\end{align}
which holds for $h=\frac{\kappa^2}{2}$ and
\begin{align}
  Q &= \frac12\int\d r\sin(\vartheta)\vartheta' = -1,\\
  \Omega &= \frac{\kappa}{2}\int\d r\,\p_r(r\sin\vartheta) = 2,
  \label{eq:Omega}
\end{align}
which are the topological charge and a boundary term,
respectively, where $\vartheta(0)=\pi$ and $\vartheta(\infty)=0$.
The Bogomol'nyi completion can also be understood as the BPS limit of
the gauged $\mathbb{C}P^1$ model with a particular choice of constant
gauge field $A_i^a = -\kappa\delta_i^a$, that generates the DM term
\cite{barton2020magnetic}.
The energy \eqref{eq:E_Bogomolnyi} shows the Bogomol'nyi or BPS
equation
\beq
\vartheta' - \frac{\sin\vartheta}{r} + \kappa(1-\cos\vartheta) = 0,
\label{eq:BPS}
\eeq
as well as the topological energy bound
\beq
E \geq 4\pi(Q + \Omega) = 4\pi,
\eeq
with equality that holds only when the BPS equation is satisfied.
The BPS equation is solved by
\beq
\vartheta = 2\arctan\left(\frac{2}{\kappa r}\right).
\eeq
This is the $\Og(3)$ sigma model lump solution with lump size
$a=\frac{2}{\kappa}$.
We also notice that the BPS bound on the energy is $4\pi$, just like
the lump energy.
Finally, we have used the asymptotic behavior of the solution
$\vartheta\sim\frac{4}{\kappa r}$ at $r\to\infty$ in evaluating the
boundary term $\Omega$ in Eq.~\eqref{eq:Omega}.

We note that the energy $\widetilde{E}$ of Eq.~\eqref{eq:En_IBP}, which
has the vorticity subtracted off, leads to a different Bogomol'nyi
bound \cite{kuchkin2020magnetic}:
\beq
\widetilde{E} \geq -4\pi.
\eeq

\subsection{Analytic solution at any coupling}\label{sec:any_coupling}

We note that the Bogomol'nyi completion requires us to select
the critical coupling $h=\frac{\kappa^2}{2}$ enabling us to complete
the square in the energy functional.
This choice of coupling corresponds to $\epsilon=2$.
However, another interesting fact about the Zeeman squared potential
($p=2$) is that inserting the Ansatz $\vartheta=2\arctan\tfrac{a}{r}$
into the full second-order Euler-Lagrange equation gives \cite{Doring2017,barton2020magnetic,ross2021skyrmion}
\beq
\frac{8a^2r(\kappa-a h)}{(a^2+r^2)^2} = 0,
\eeq
which beautifully has the solution $a=\frac{\kappa}{h}=\ell$.
This size does not contradict the BPS solution, since the first-order
BPS equation required $h=\frac{\kappa^2}{2}$ which in the full
second-order solution corresponds to the lump size
$a = \frac{\kappa}{h} = \frac{2}{\kappa}$.
It is interesting, however to insert the exact solution into the
energy functional \eqref{energy-integral2D}, which gives \cite{barton2020magnetic,kuchkin2020magnetic}
\beq
E = 2\pi\int r\d r\,\frac{4h^2\kappa^2}{(h^2r^2+\kappa^2)^2} = 4\pi.
\eeq
But this is nothing but the lump energy.
We see that the $p=2$ case is a rather special case, where the energy
is always the lump energy and the solution is always the lump
solution with the lump size $a=\ell=\frac{\kappa}{h}$. However, only at the
critical coupling, the second-order equation of motion reduces to the
first-order BPS equation \eqref{eq:BPS}.\footnote{The fact that this
analytic solution holds for all values of the coupling, is somewhat
similar to the situation in the baby-Skyrme model
for $p=4$ \cite{Leese:1989gi,Bolognesi:2014ova}.} 

The $p=2$ case is thus applicable to many more physical systems, than
just the critically coupled one \cite{Doring2017,barton2020magnetic,ross2021skyrmion}.
Having the analytic solution in hand may be advantageous for further
studies, including interaction and scattering etc.

Using the energy functional $\widetilde{E}$ of Eq.~\eqref{eq:En_IBP},
where the vorticity is subtracted off, the energy becomes instead
\begin{align}
\widetilde{E} &= 2\pi\int r\d r\;\frac{4h\kappa^2(h-\kappa^2)}{(h^2r^2+\kappa^2)^2}\non
&=4\pi(1-\epsilon).
\end{align}
We may interpret the change of sign of the energy at
$\epsilon=\epsilon^{\rm crit}=1$ as the Skyrmionic phase being
energetically preferable for $\epsilon>1$ and the homogeneous phase on
the other hand preferable for $\epsilon<1$.
Luckily, the BPS-case resides in the Skyrmionic phase at $\epsilon=2$.

Since the $p=2$ case is an exact solution for all values of
$\epsilon$, there is no need to perform numerical computations for
this case.

\subsection{Numerical solutions}\label{sec:numerical}

For the general theory, we can solve equation \eqref{profile equation 2D}
using a standard shooting method, beginning from the origin with the
initial condition $\vartheta(0) = \pi$.
The form of the solution close to the origin is
$\vartheta(r) \simeq \pi - \gamma r$.
By varying the angular coefficient $\gamma$, we find only one solution in
the topological sector $Q=-1$ as already shown in the literature
\cite{melcher2014chiral, schroers2019gauged, kuchkin2020magnetic,
  barton2020magnetic, walton2020geometry, schroers2021solvable,
  ross2021skyrmion}.

\begin{figure*}[!htp]
  \centering
  \mbox{\subfloat[$p=\tfrac32$]{\includegraphics[width=0.49\textwidth]{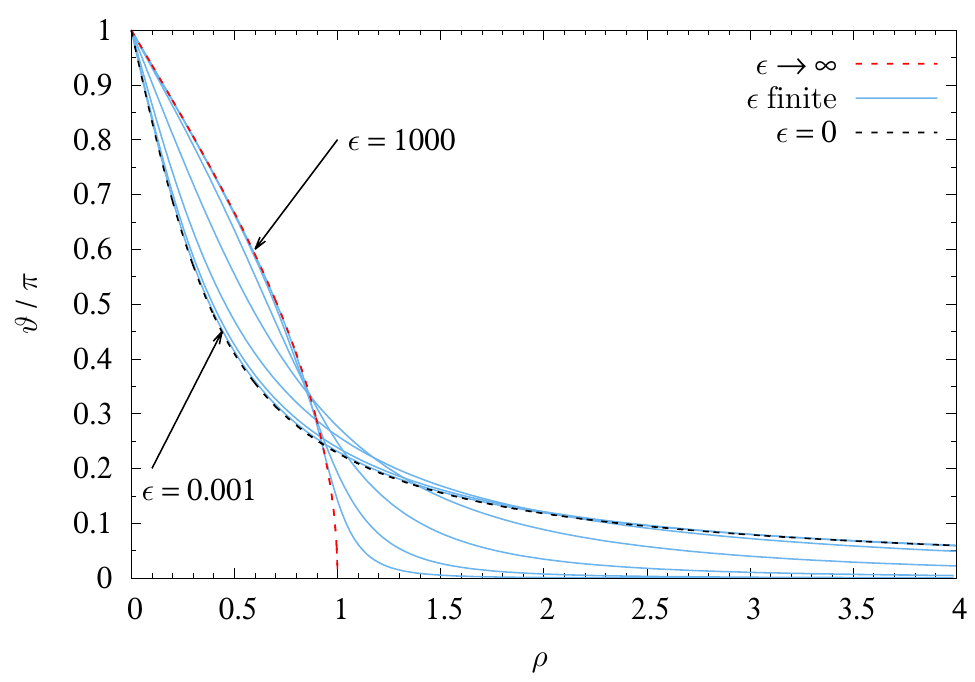}}
    \subfloat[$p=1$]{\includegraphics[width=0.49\textwidth]{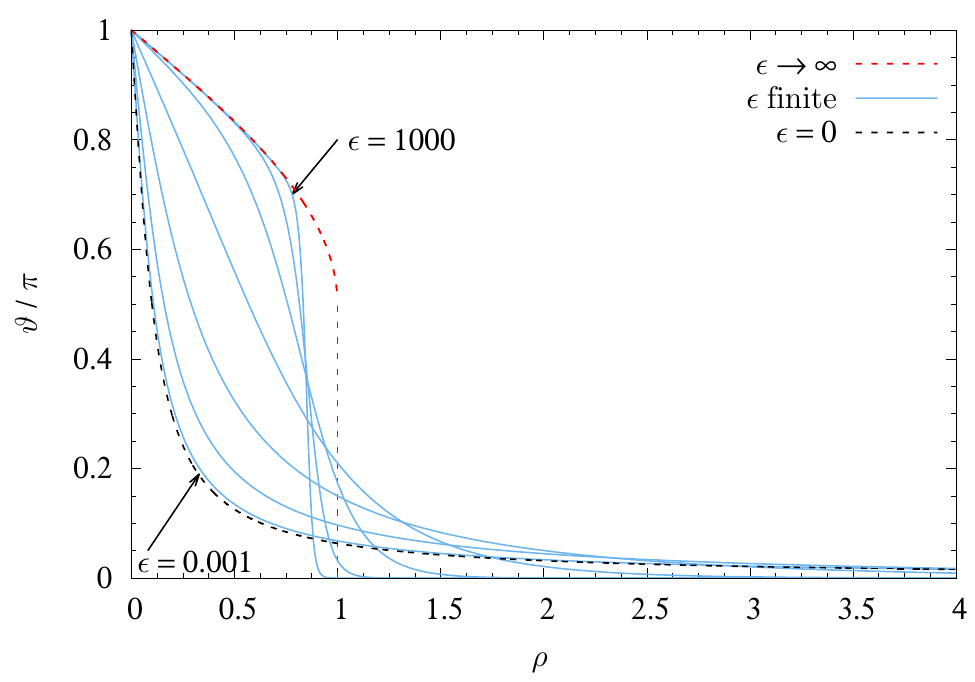}}}
  \caption{Profile function $\vartheta(\rho)$ for (a) $p=\frac{3}{2}$ and
    (b) $p=1$ with $\epsilon=0.001$, $0.01$, $0.1$, $1$, $10$, $100$, $1000$.
    The red-dashed curve corresponds to the restricted limit, whereas
    the black-dashed curve is the sigma-model lump limit with lump
    sizes (a) $a=\frac{\ell}{\sqrt{2}}$ and (b) $a=\ell/5$.
    The latter is chosen for illustrative purposes.
  }
  \label{fig:profiles}
\end{figure*}

\begin{figure*}[!htp]
  \centering
  \mbox{\subfloat[$p=\tfrac32$]{\includegraphics[width=0.49\textwidth]{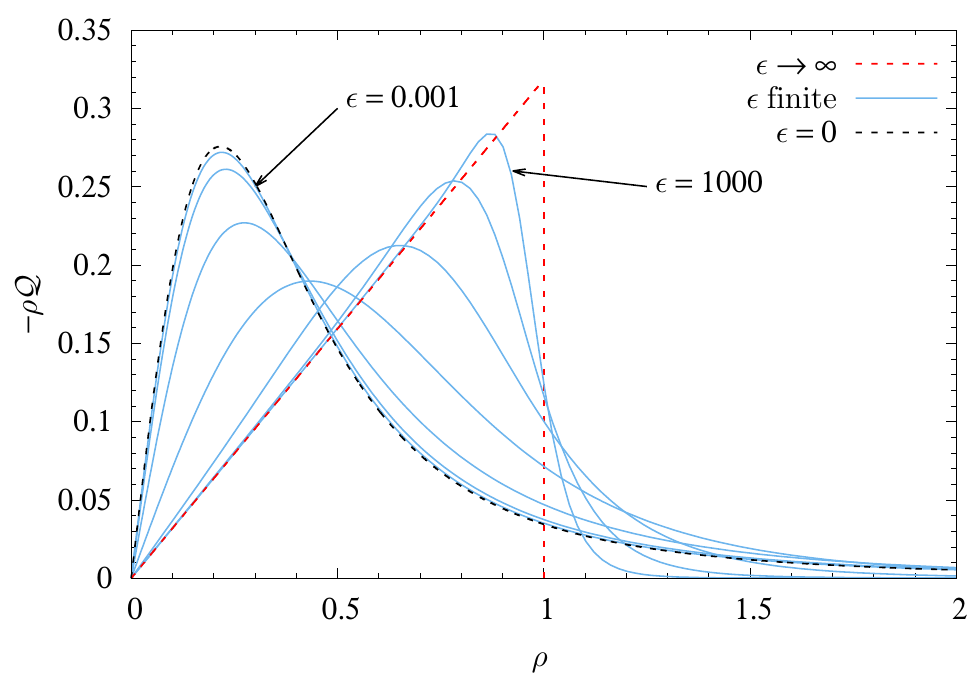}}
    \subfloat[$p=1$]{\includegraphics[width=0.49\textwidth]{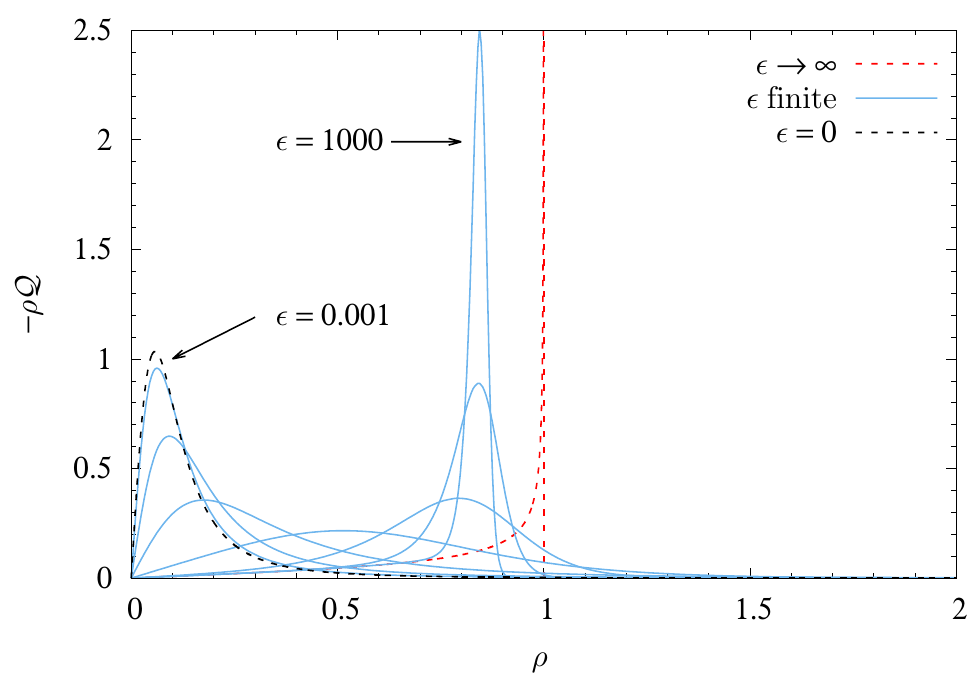}}}
  \caption{Topological charge density,
    $\calQ$
    for (a) $p=\frac32$ and (b) $p=1$ with $\epsilon=0.001$, $0.01$,
    $0.1$, $1$, $10$, $100$, $1000$.
    The topological charge densities of the lumps and restricted
    Skyrmions are shown with dashed black and red curves, respectively.
    In the $p=1$ case, the size of the lump is selected for
    illustrative purpose, corresponding to the solutions in
    Fig.~\ref{fig:profiles}.
  }
  \label{fig:topocharge}
\end{figure*}

Let us now study how the theory behaves by varying the control
parameters, $\ell$ and $\epsilon$.
We plot the solutions $\vartheta$ for a variety of couplings,
$\epsilon$, for $p=\frac32$ (panel (a)) and $p=1$ (panel (b)) in
Fig.~\ref{fig:profiles}.
In each case, we plot $\vartheta$ as a function of $\rho$ which is
proportional to $r/\ell$, such that $\ell$-dependence is removed from
the solutions and $\rho$ is defined in
Eqs.~\eqref{restrp1}-\eqref{restrp3/2}. 
We see that as $\epsilon$ goes from $0$ to $\infty$ the solution flows
from the sigma model lump solutions (black dashed lines) to the
restricted (compacton) solution (red dashed lines) in both cases.
In the $p=\frac32$ case, the lump solution \eqref{profilelump} with a
lump size \eqref{eq:lumpsize_a} is shown and the restricted solution is
that of Eq.~\eqref{restrp3/2}.
We notice that the numerical solutions tend to the lump solution with
the correct lump size \eqref{eq:lumpsize_a}, where a boundary term has
been discarded when determining the lump size from the energy
functional.
In the $p=1$ case, on the other hand, we have set the lump size to
$a=\frac{\ell}{5}$ by hand for visualization purposes and the
restricted supercompacton solution that the solutions flow to for
large $\epsilon$ is given in Eq.~\eqref{restrp1}.
We see again that the lump solution matches well with the numerical
solutions, but the lump size is set by hand for comparison
with the solution at small $\epsilon=10^{-3}$.
We will shortly argue that the lump size should go to zero in the
limit $\epsilon\to0$.
Note that the supercompacton has a discontinuity and covers only half 
of the 2-sphere, from the south pole to the equator.
At finite $\epsilon$, the discontinuity is resolved and topological
soliton covers the entire sphere.
As $\epsilon\to\infty$, a step-function type discontinuity is created 
and the missing half of the sphere is mapped into a very narrow region. 

Fig.~\ref{fig:topocharge} shows the topological charge density that
integrates to $Q=-1$, in the $p=\frac32$ case (panel (a)) and in the $p=1$ case (panel
(b)).

\begin{figure*}[!htp]
  \centering
  \mbox{\subfloat[$p=\tfrac32$]{\includegraphics[width=0.49\textwidth]{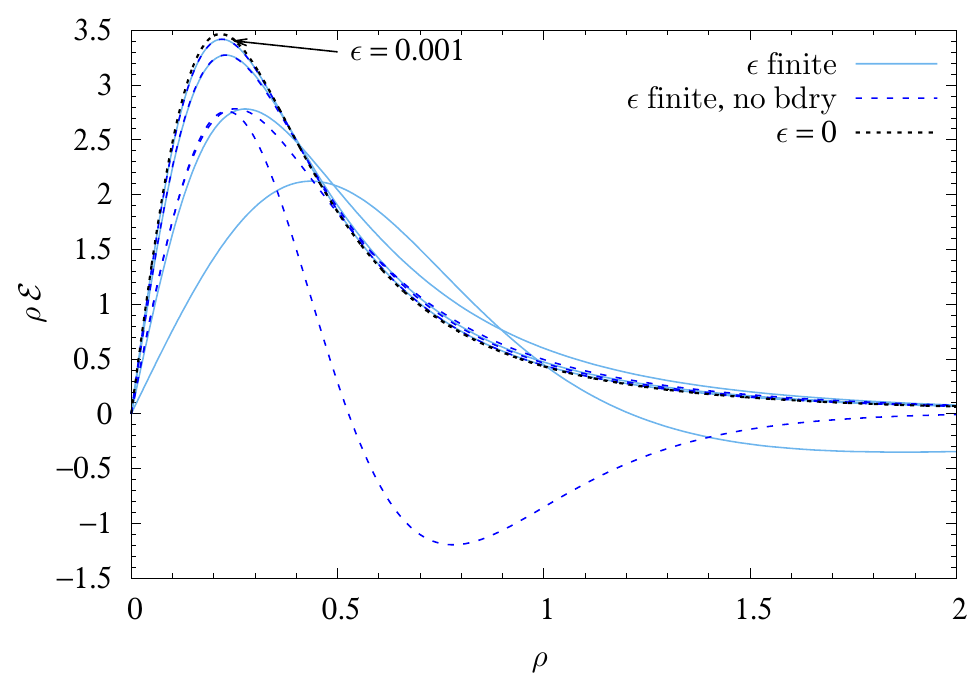}}
    \subfloat[$p=1$]{\includegraphics[width=0.49\textwidth]{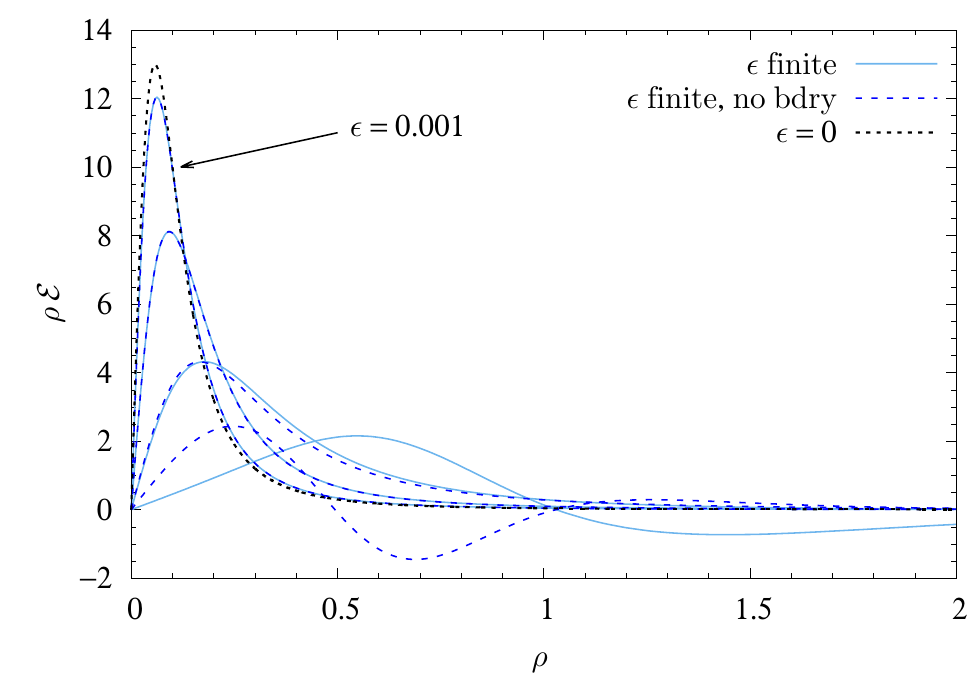}}}
  \mbox{\subfloat[$p=\tfrac32$]{\includegraphics[width=0.49\textwidth]{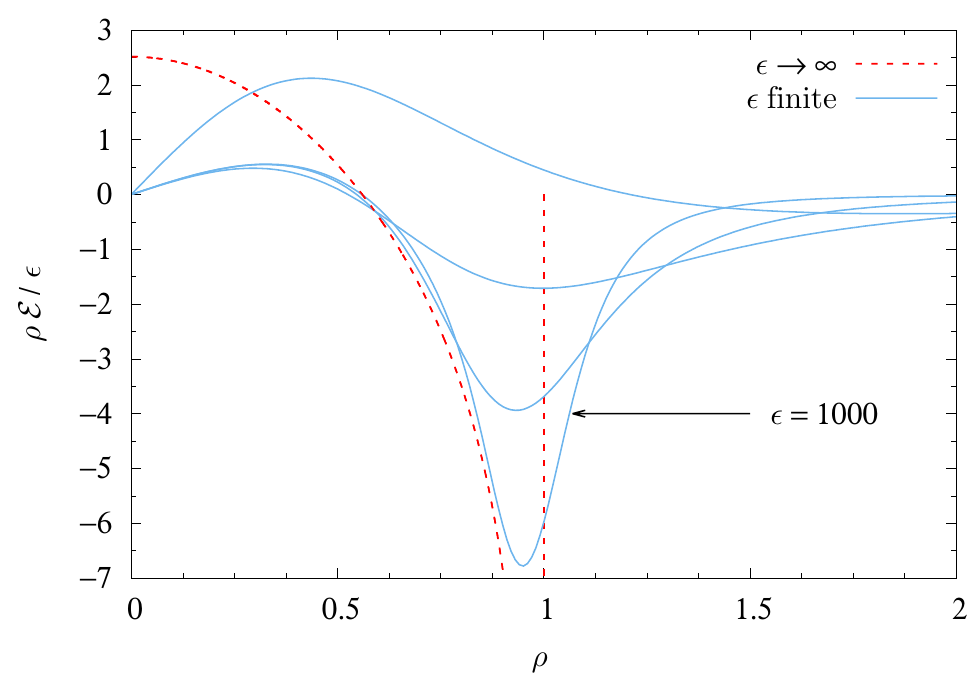}}
    \subfloat[$p=1$]{\includegraphics[width=0.49\textwidth]{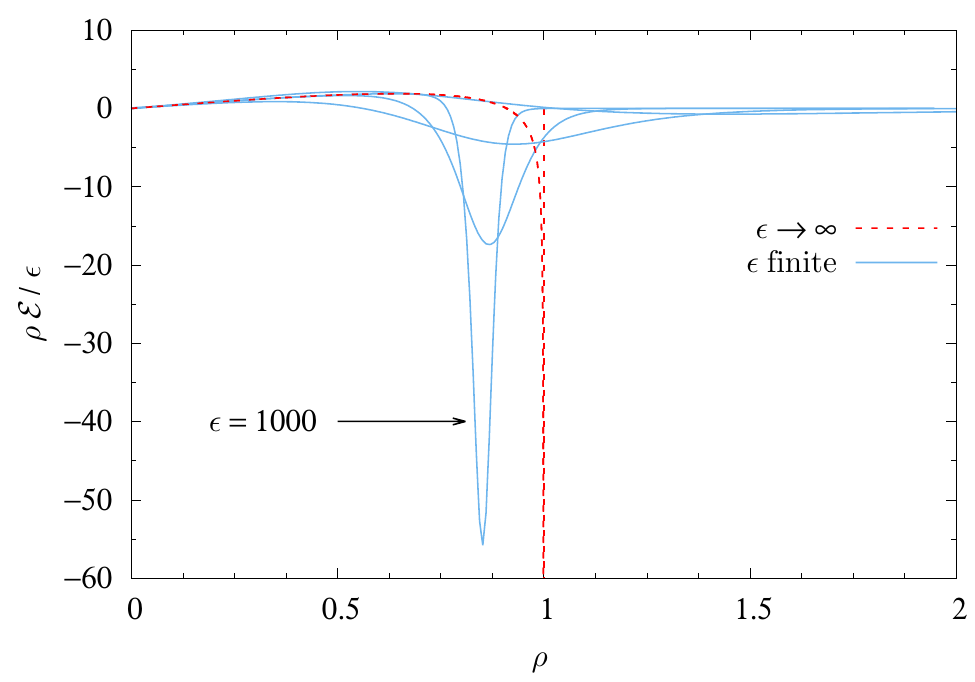}}}
  \mbox{\subfloat[$p=\tfrac32$]{\includegraphics[width=0.49\textwidth]{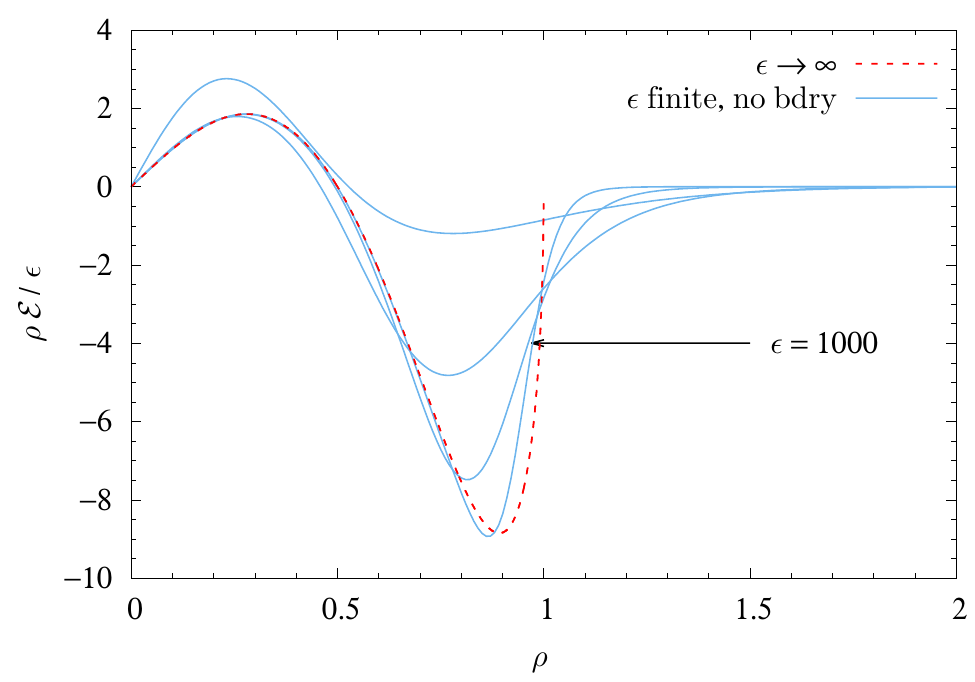}}
    \subfloat[$p=1$]{\includegraphics[width=0.49\textwidth]{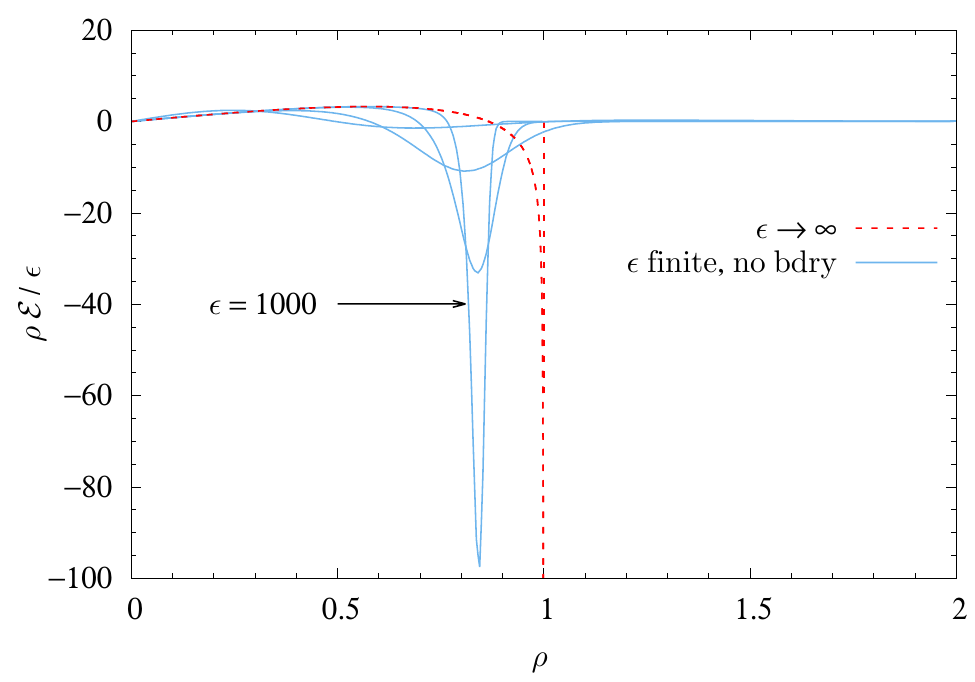}}}
  \caption{Energy density: 
    (a) $p=\frac32$ (b) $p=1$ for $\epsilon=0.001,0.01,0.1,1$
    (solid blue lines) and with the boundary term
    \eqref{eq:Omegatilde} subtracted off (dashed dark-blue lines).
    Energy density: (c) $p=\frac32$ (d) $p=1$ for $\epsilon =1,10,100,1000$
    and finally (e) $p=\frac32$ (f) $p=1$ and $\epsilon=1,10,100,1000$
    with the boundary term \eqref{eq:Omegatilde} subtracted off.
    }
  \label{fig:energies}
\end{figure*}

Fig.~\ref{fig:energies} shows the energy densities for the
sigma-model lump limit ($\epsilon\to0$) in panels (a) and (b)
with and without the boundary term for
$p=\frac32$ and $p=1$, respectively, and for the restricted limit
($\epsilon\to\infty$) in panels (c) and (d) for $p=\frac32$ and $p=1$, respectively,
in panels (e) and (f) $p=\frac32$ and
$p=1$, respectively, with the boundary term \eqref{eq:Omegatilde}
subtracted off.
In both the charge density as well as in the energy density, the
discontinuity that develops in the $p=1$ case for $\epsilon\to\infty$
is clearly seen in Figs.~\ref{fig:topocharge}(b) and
\ref{fig:energies}(d), respectively.
We note that the energy density is divergent (singular) in the $p=\frac32$ case
at $\rho=1$ when the boundary term is taken into account, see
Fig.~\ref{fig:energies}(c), but is convergent (finite) when the
boundary term is dropped, see Fig.~\ref{fig:energies}(e).
We also note that the total energy (the integral) is unchanged in
the $p=1$ case (see Figs.~\ref{fig:energies}(d) and
\ref{fig:energies}(f)), but the energy density is slightly modified
by the presence of the boundary term.
  
\begin{figure*}[!htp]
  \centering
  \mbox{
    \subfloat[]{\includegraphics[width=0.49\linewidth]{{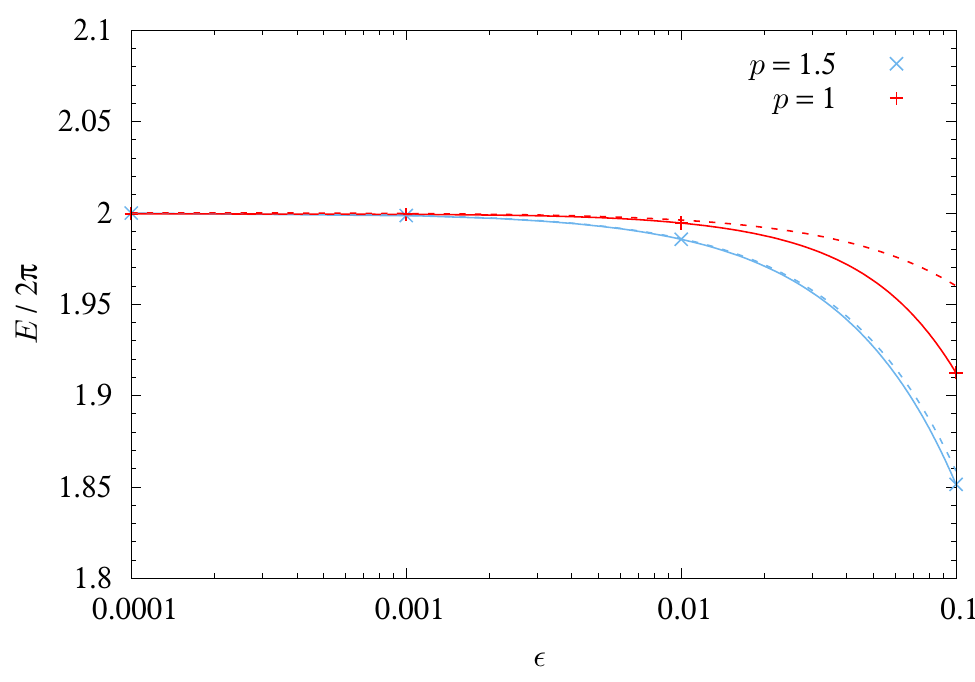}}}
    \subfloat[]{\includegraphics[width=0.49\linewidth]{{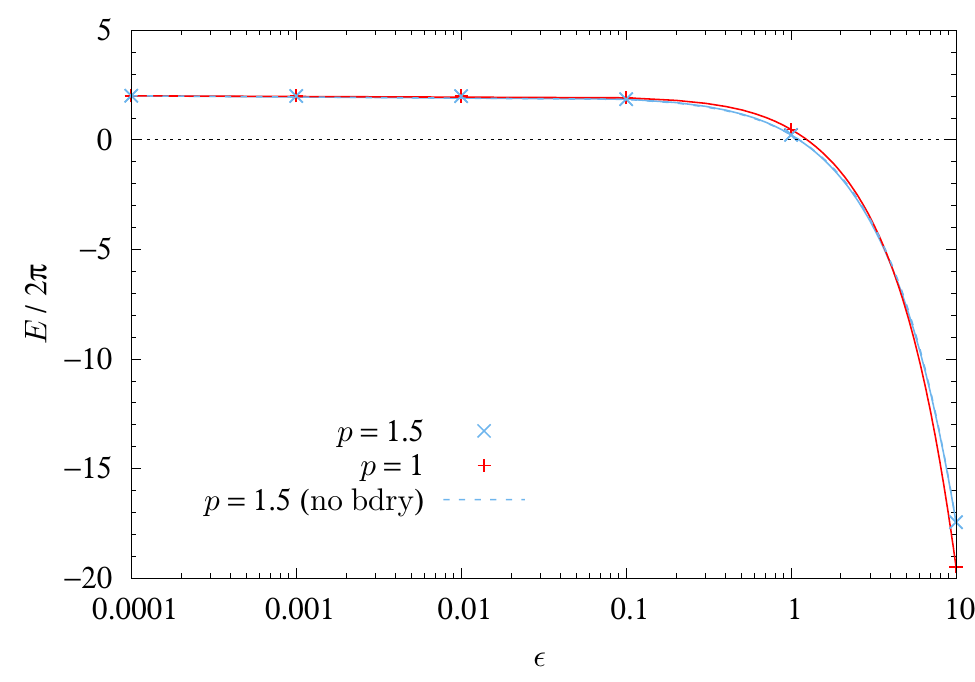}}}}
  \caption{Energy in the lump limit ($\epsilon\to 0$) as a function of
    $\epsilon$ for $p=\tfrac32$ and $p=1$.
    (a) The blue dashed line corresponds to the analytical energy formula \eqref{eq:Ea_sol} for
    the lump in the limit of small but nonvanishing $\epsilon$
    in the case of $p=\frac32$.
    The red dashed line instead is the energy of Eq.~\eqref{eq:Ep=1reg} with the
    lump size taken from Fig.~\ref{fig:profiles}.
    (b) The energy shown up to larger values of $\epsilon$ revealing
    the sign change near $1<\epsilon<2$.
    The case $p=\frac32$ is shown for the full energy (solid blue
    line) and with the boundary term \eqref{eq:Omegatilde} subtracted
    off (dashed blue line).
  }
  \label{fig:entab}
\end{figure*}

We show the total energy of the numerical solutions in the
lump limit ($\epsilon\to0$) for the $p=\frac32,1$ cases in
Fig.~\ref{fig:entab}(a).
We notice that the $p=\frac32$ case is well described by the energy
functional \eqref{eq:Ea_sol} with the correct lump size
\eqref{eq:lumpsize_a}.
This confirms that discarding the boundary term \eqref{eq:Omegatilde}
is appropriate for computing the energies in the lump limit,
which is also the case where the variational problem is well
defined \cite{kuchkin2020magnetic}.
In the $p=1$ case, we show the regularized energy \eqref{eq:Ea_cutoff}
with a dashed red curve and lump size $a$ as chosen in
Fig.~\ref{fig:profiles}.
We show the sign change of the energy in the $p=\frac32,1$
case in Fig.~\ref{fig:entab}(b) which appears to take place around
$1<\epsilon<2$, which is consistent with the lump prediction
\eqref{eq:Ea_sol} which for $p=\frac32$ should take place at
$\epsilon=\epsilon^{\rm crit}=\sqrt{2}$.

\begin{figure}[!htp]
  \centering
  \includegraphics[width=0.49\linewidth]{{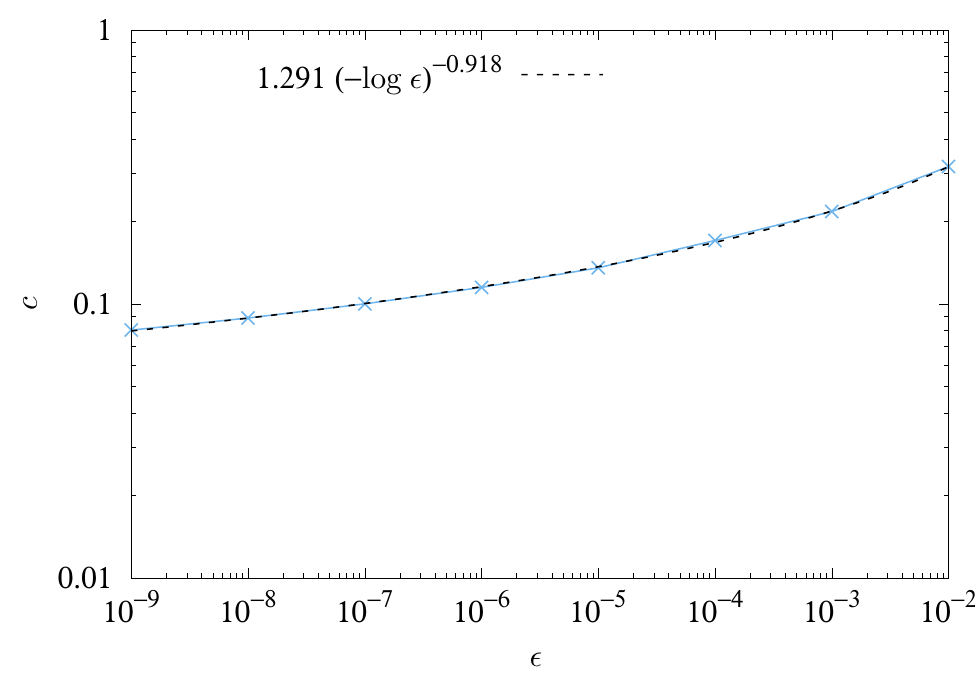}}
  \caption{
    The lump size $c=\frac{a}{\ell}$ estimated by fitting the lump
    solution to the numerical solutions.
    The dashed black curve is a logarithmic fit.
  }
  \label{fig:epsilonto0}
\end{figure}

Finally, we consider the difficult problem of estimating the lump size
in the $\epsilon\to 0$ limit.
Since we can see from Fig.~\ref{fig:profiles} that the profiles match
the lump solution if the lump size is fitted correctly, we can extract
this lump size from the numerical solutions by fitting the
size of the lump solution to the numerical ones.
Fig.~\ref{fig:epsilonto0} shows the numerically extracted
lump sizes and a logarithmic fit finding an approximate behavior of
the lump size as $c\approx 1.291(-\log\epsilon)^{-0.918}$.
Clearly the lump size goes very slowly to zero.

\subsection{Comparison with experiments}\label{sec:exp}

We have confirmed that the magnetic Skyrmion model, up to rescaling of
energy and length, is essentially a one-parameter model \eqref{derrick1}-\eqref{derrick2}.
This parameter, $\epsilon$, controls the relation between $E_2$ and
$E_1+E_0$.
In both the large-$\epsilon$ and the small-$\epsilon$ limit, analytic
approximations are available, see Secs.~\ref{sec:lump} and
\ref{sec:restricted}. 
It may be interesting to explore various phenomenological realizations
of chiral magnets and determine the order of magnitude of the
parameter $\epsilon$ that can be realized.

The typical Skyrmion size, determined by the DM interaction, in chiral
magnets, e.g.~MnSi \cite{muhlbauer2009skyrmion, nagaosa2013topological}, is about
$\ell \simeq 5 - 100\nm$.
From the phase diagram for MnSi \cite{fert2013skyrmions}, the range for the external magnetic
field in the Skyrmion phase is $B \sim 0.1 - 0.3\Tesla$.
Another important experimental parameter is $\lambda$, the period of
the spiral, determined by the ratio of the dipolar and exchange
interactions: 
\begin{equation}\label{lambda-period}
  \lambda = \dfrac{2\pi J}{D} .
\end{equation}
Such parameter is well known for different chiral magnets (see e.g.~Table 1
in Ref.~\cite{nagaosa2013topological}), in the case of MnSi:
$\lambda=18\nm$.
From Eq.~\eqref{sizes}, we get 
\begin{equation}
  \begin{split}
    J &= \frac{\lambda D}{2\pi} = \frac{\lambda \ell |B|}{4\pi} ,\\
    \epsilon &= \dfrac{4 D^2}{J|B|} = 4\pi \dfrac{\ell}{\lambda}.
  \end{split}
\end{equation}
Using the minimal values for the previous parameters (for MnSi)
defined on a range of values,
i.e.~$B \sim 10^{-1}\Tesla, \ell \sim 10\nm$, and $\lambda \simeq 10\nm$, we find 
\begin{equation}
  \begin{split}
    J &\simeq 100\nm^2 \times\Tesla \simeq 10^{-14}\Joule,\\
    \epsilon &\simeq 10,
  \end{split}
\end{equation}
where $\Joule$ is the unit \emph{Joule}, not to be confused with the
interaction strength of the Heisenberg Hamiltonian.
Note that in general $\ell$ can change significantly, viz.~one order
of magnitude.
For instance, if the we fix $\ell = 100\nm$ and
$\lambda\in\mathcal{O}(10^1)$ nm as before, we find that $\epsilon$
changes one order of magnitude, i.e.~$\epsilon \simeq 100$.
For materials other than MnSi, e.g.~Fe$_{1-x}$Co$_x$Si or
Mn$_{1-x}$Fe$_x$Ge, $\lambda$ can also be of the order $\mathcal{O}(10^2)\nm$.
Therefore, considering the other two cases where $\lambda$ is fixed as
$\lambda\sim 10^2\nm$ and $\ell$ takes its maximum and minimum values, we find
the relevant ranges for $\epsilon$:
\begin{equation}\label{exper-values}
  \begin{split}
    1 &\lesssim \epsilon \lesssim 10, \qquad \ell \simeq 10\nm,\\
    10 &\lesssim \epsilon \lesssim 100, \quad\;\, \ell \simeq 100\nm.
  \end{split}
\end{equation}
Finally, the approximating limits studied in this work can be seen to
be relevant for describing certain experimental 2D magnetic Skyrmions within the ranges given in the above equation.

In Secs.~\ref{sec:critical} and \ref{sec:any_coupling}, we
saw that in the special physical case of $p=2$, which
describes chiral magnets with anisotropy interactions, any spherical
magnetic Skyrmion can be represented by the analytical lump solution,
regardless of its DM coupling, $\kappa$.
However, the potential $V_{p=2}$ corresponds to the sum of a magnetic
field and anisotropy (easy-plane) contributions.
To realize this potential, the couplings of the anisotropy
term must be exactly half of that of the Zeeman term.
It turns out to be possible \cite{Banerjee:2014hna}.
Furthermore, since the critical coupling regime corresponds to the
condition $\forall~\ell, \ \epsilon = 2$, which satisfies the
inequalities \eqref{exper-values}, we suggest that, in principle, a
BPS magnetic Skyrmion could be realized experimentally. 

\section{Conclusion}\label{sec:conclusion}

In this paper, we have studied magnetic Skyrmions in chiral magnets
with three different potentials, corresponding to $p=1$, $p=\frac32$
and $p=2$ in the limits of the control parameter
$\epsilon=\frac{\kappa^2}{h}=\frac{4D^2}{|B|J}$ becoming very large and
going to zero.
The $p=1$ case is simply the Zeeman potential term and the
$p=2$ case is a particular combination of the easy-plane
(anisotropy) and the Zeeman potentials.
The $p=\frac32$ case is probably somewhat more academic.
The $\epsilon$ vanishing limit reduces the Skyrmion to a sigma model lump with a
lump size that we accurately estimated, except for the case of the
Zeeman term ($p=1$).
The large-$\epsilon$ limit instead corresponds to the “restricted"
magnetic Skyrmions, for which we found the exact analytic solutions as
well.
In the $p=\frac32$ case, the restricted solution is a compacton,
similar in nature to those found in the baby-Skyrme \cite{Gisiger:1996vb,Adam:2009px,Adam:2010jr}
and 3D Skyrme models \cite{Adam:2010fg}.
However, in the $p=1$ case corresponding to the Zeeman term, we find a
\emph{new} type of restricted solution, which we dub a \emph{supercompacton}, since
it becomes discontinuous.
We provide numerical solutions showing that they slowly tend to the
supercompacton limit for very large values of $\epsilon$.
We have also reviewed the beautiful result of the critically coupled
magnetic Skyrmion of Barton-Singer-Ross-Schroers
\cite{barton2020magnetic}, where an analytic solution is available,
hence providing many studies with a clear advantage.
Their solution is known to be an exact solution for any value of the
coupling $\epsilon$ \cite{Doring2017,barton2020magnetic,ross2021skyrmion}, but we find here that
dropping the boundary term that leads to the correct lump size for
this solution also changes the energy from the lump energy for all
values of $\ell$ to the energy functional $4\pi(1-\epsilon)$.
This energy functional changes sign at $\epsilon=1$ suggesting
$\epsilon=1$ being the critical coupling for having the Skyrmionic
phase.
We finally emphasize that our study, within the defined
restricted limit, provides analytical solutions for magnetic
Skyrmions in certain magnetic material, as shown in Sec.~\ref{sec:exp}.

Future directions that would be interesting to study, would be to
utilize the exact analytic solutions for dynamical problems such as
acceleration, interaction and scattering.
Extending the planar materials to 3D materials may also be
interesting, where numerical computations have found braiding magnetic
Skyrmion solutions \cite{Zheng2021}. Using various limits, perhaps
analytic insight can be found also in such cases.

\section*{Acknowledgments}
We thank Bruno Barton-Singer and Calum Ross for comments on our draft.
The work of S.~B. is supported by the INFN special research project
grant ``GAST'' (Gauge and String Theories).
S.~B.~G. thanks the Outstanding Talent Program of Henan University and
the Ministry of Education of Henan Province for partial support.
The work of S.~B.~G.~is supported by the National Natural Science
Foundation of China (Grant No.~12071111) and by the Ministry of
Science and Technology of China (Grant No.~G2022026021L).
R.~M. thanks the Department of Physics ``E.~Fermi'' of University of
Pisa where this work was initiated.

\appendix
\section{Continuum Limit}\label{appendix}
The Heisenberg term of the GL energy density can be derived from the
continuum limit of the lattice Hamiltonian 
\begin{equation}\label{Heisenberg_discrete}
    \mathcal{H}_{\rm H} = - J \sum_{i, \mu} \bm{n}_i \cdot \bm{n}_{i + a \hat{\bm{e}}_{\mu}} ,
\end{equation}
where $a \hat{\bm{e}}_{\mu}$ is the vector connecting the $i$-lattice
site with its neighboring sites $i + a \hat{\bm{e}}_{\mu}$, on a
square ($\mu=x,y$) lattice. Assuming a small lattice constant $a$, one
can perform a Taylor expansion on $\bm{n}_{i + a \hat{\bm{e}}_{\mu}}$,
i.e.,
\begin{equation}
    \bm{n}_{i + a \hat{\bm{e}}_{\mu}} \simeq \bm{n}_i + a \partial_{\mu} \bm{n}_i + (a^2/2) \partial^2_{\mu} \bm{n}_i + \cdots
\end{equation}
and insert it into Eq.~\eqref{Heisenberg_discrete}.
One finds that first term $\bm{n}_i \cdot \bm{n}_i$ is equal to the
unity and can be discarded as a constant term.
The second term, proportional to $\bm{n}_i\cdot\partial_{\mu}\bm{n}_i$,
vanishes due to nonlinear constraint $\bm{n}\cdot\bm{n}=1$.
However, the third term
\begin{equation}
    \mathcal{H}_{\rm H} \approx - \dfrac{1}{2}Ja^2 \sum_i \bm{n}_i \cdot \partial^2_{\mu} \bm{n}_i ,
\end{equation}
in the continuum limit, $a \to 0$, and after an integration by parts,
leads to a non-vanishing contribution, i.e., 
\begin{equation}
    E_2[\bm{n}] = \dfrac{1}{2}J \int\d^2x \  (\partial_{\mu} \bm{n} \cdot \partial_{\mu} \bm{n}) , \qquad \bm{n}\cdot\bm{n}=1 .
\end{equation}
The DM interaction, at the lattice level can be written as 
\begin{equation}
    \mathcal{H}_{\rm DM} = - D \sum_{i,\mu} \hat{\bm{e}}_{\mu} \cdot (\bm{n}_i \times \bm{n}_{i + a \hat{\bm{e}}_{\mu}}) ,
\end{equation}
where we expressed the DM vector as $\bm{D} = D\hat{\bm{e}}_{\mu}$.
Upon Taylor expansion, the first derivative of
$\bm{n}_{i  + a \hat{\bm{e}}_{\mu}}$, provides the non-null
contribution 
\begin{equation}
    \mathcal{H}_{\rm DM} \approx D a \sum_{i,\mu}\bm{n}_i \cdot (\hat{\bm{e}}_{\mu} \times  \partial_{\mu} \bm{n}_i) = Da \sum_i \bm{n}_i \cdot (\nabla \times \bm{n}_i),
\end{equation}
where at the second step we have applied the vector definition for the
curl operator. Thus, in the continuum limit, one finds 
\begin{equation}\label{dm}
    E_1[\bm{n}] = D a^{-1} \int\d^2x \ \bm{n} \cdot (\nabla \times \bm{n}) ,
\end{equation}
which corresponds to the usual DM energy functional. Finally, let us
consider the potential-like term $- \bm{B} \cdot \bm{n}$ that
describes the Zeeman interaction with an external magnetic field,
which we choose to be aligned along the third direction,
$\bm{B} = B\bm{e}_3$. Since we have arbitrarily chosen the vacuum field
$\bm{N}=\bm{e}_3$, and we need the potential to vanish when evaluated on
the vacuum, the energy contribution can be written as
$E_0[\bm{n}]=B(1 - \bm{N}\cdot \bm{n}) = B(1-n_3)$.
Such potential contribution can
be generalized to a class of potentials $V_p$ of the
form~\eqref{generic-potential}. At this point, by considering all the
three contributions to the whole energy functional $E[\bm{n}]$ one
gets Eq.~\eqref{energy-functional2D-RIGHT}.

\bibliographystyle{JHEP}
\bibliography{biblio}

\providecommand{\href}[2]{#2}\begingroup\raggedright\begin{thebibliography}{10}

\bibitem{rajaraman1989introduction}
R.~Rajaraman, \emph{Solitons and Instantons}, vol.~15. Elsevier, Amsterdam,
  1st~ed., 1987.

\bibitem{MantonSutcliffe}
N.~Manton and P.~Sutcliffe, \emph{Topological Solitons}, Cambridge Monographs
  on Mathematical Physics. Cambridge University Press, 2004,
  \href{https://doi.org/10.1017/CBO9780511617034}{10.1017/CBO9780511617034}.

\bibitem{Braun2012}
H.-B. Braun, \emph{Topological effects in nanomagnetism: from
  superparamagnetism to chiral quantum solitons},
  \href{https://doi.org/10.1080/00018732.2012.663070}{\emph{Advances in
  Physics} {\bfseries 61} (2012) 1}.

\bibitem{shifman2012advanced}
M.~Shifman, \emph{Advanced Topics in Quantum Field Theory: A Lecture Course}.
  Cambridge University Press, 2012,
  \href{https://doi.org/10.1017/CBO9781139013352}{10.1017/CBO9781139013352}.

\bibitem{Shnir_2018}
Y.~M. Shnir, \emph{Topological and Non-Topological Solitons in Scalar Field
  Theories}, Cambridge Monographs on Mathematical Physics. Cambridge University
  Press, 2018,
  \href{https://doi.org/10.1017/9781108555623}{10.1017/9781108555623}.

\bibitem{han2017skyrmions}
J.~H. Han, \emph{Skyrmions in condensed matter}, vol.~278. Springer, 2017,
  \href{https://doi.org/10.1007/978-3-319-69246-3}{10.1007/978-3-319-69246-3}.

\bibitem{skyrme1961non}
T.~H.~R. Skyrme, \emph{A non-linear field theory},
  \href{https://doi.org/10.1098/rspa.1961.0018}{\emph{Proceedings of the Royal
  Society of London. Series A. Mathematical and Physical Sciences} {\bfseries
  260} (1961) 127}.

\bibitem{skyrme1962unified}
T.~H.~R. Skyrme, \emph{A unified field theory of mesons and baryons},
  \href{https://doi.org/10.1016/0029-5582(62)90775-7}{\emph{Nuclear Physics}
  {\bfseries 31} (1962) 556}.

\bibitem{belavin1975i}
A.~Belavin and A.~Polyakov, \emph{Metastable states of two-dimensional
  isotropic ferromagnets}, {\emph{JETP letters} {\bfseries 22} (1975) 245}.

\bibitem{bogdanov1989thermodynamically}
A.~N. Bogdanov and D.~Yablonskii, \emph{Thermodynamically stable “vortices”
  in magnetically ordered crystals. the mixed state of magnets}, {\emph{JETP}
  {\bfseries 68} (1989) 101}.

\bibitem{bogdanov1994thermodynamically}
A.~Bogdanov and A.~Hubert, \emph{Thermodynamically stable magnetic vortex
  states in magnetic crystals}, {\emph{Journal of Magnetism and Magnetic
  Materials} {\bfseries 138} (1994) 255}.

\bibitem{roessler2006spontaneous}
U.~K. R{\"o}{\ss}ler, A.~N. Bogdanov and C.~Pfleiderer, \emph{Spontaneous
  skyrmion ground states in magnetic metals},
  \href{https://doi.org/10.1038/nature05056}{\emph{Nature} {\bfseries 442}
  (2006) 797}.

\bibitem{Rossler:2010st}
U.~Rossler, A.~A. Leonov and A.~N. Bogdanov, \emph{{Chiral Skyrmionic matter in
  non-centrosymmetric magnets}},
  \href{https://doi.org/10.1088/1742-6596/303/1/012105}{\emph{J. Phys. Conf.
  Ser.} {\bfseries 303} (2011) 012105}
  [\href{https://arxiv.org/abs/1009.4849}{{\ttfamily 1009.4849}}].

\bibitem{Ezawa:2010uy}
M.~Ezawa, \emph{{Compact Skyrmions, Merons and Bimerons in Thin Chiral Magnetic
  Films}}, \href{https://doi.org/10.1103/PhysRevB.83.100408}{\emph{Phys. Rev.
  B} {\bfseries 83} (2011) 100408}
  [\href{https://arxiv.org/abs/1010.4119}{{\ttfamily 1010.4119}}].

\bibitem{Banerjee:2014hna}
S.~Banerjee, J.~Rowland, O.~Erten and M.~Randeria, \emph{Enhanced stability of
  skyrmions in two-dimensional chiral magnets with rashba spin-orbit coupling},
  \href{https://doi.org/10.1103/PhysRevX.4.031045}{\emph{Phys. Rev. X}
  {\bfseries 4} (2014) 031045}.

\bibitem{melcher2014chiral}
C.~Melcher, \emph{Chiral skyrmions in the plane},
  \href{https://doi.org/10.1098/rspa.2014.0394}{\emph{Proceedings of the Royal
  Society A: Mathematical, Physical and Engineering Sciences} {\bfseries 470}
  (2014) 20140394}.

\bibitem{Rybakov:2018bxt}
F.~N. Rybakov and N.~S. Kiselev, \emph{{Chiral magnetic skyrmions with
  arbitrary topological charge}},
  \href{https://doi.org/10.1103/PhysRevB.99.064437}{\emph{Phys. Rev. B}
  {\bfseries 99} (2019) 064437}
  [\href{https://arxiv.org/abs/1806.00782}{{\ttfamily 1806.00782}}].

\bibitem{schroers2019gauged}
B.~J. Schroers, \emph{{Gauged Sigma Models and Magnetic Skyrmions}},
  \href{https://doi.org/10.21468/SciPostPhys.7.3.030}{\emph{SciPost Phys.}
  {\bfseries 7} (2019) 030} [\href{https://arxiv.org/abs/1905.06285}{{\ttfamily
  1905.06285}}].

\bibitem{barton2020magnetic}
B.~Barton-Singer, C.~Ross and B.~J. Schroers, \emph{{Magnetic Skyrmions at
  Critical Coupling}},
  \href{https://doi.org/10.1007/s00220-019-03676-1}{\emph{Commun. Math. Phys.}
  {\bfseries 375} (2020) 2259}
  [\href{https://arxiv.org/abs/1812.07268}{{\ttfamily 1812.07268}}].

\bibitem{kuchkin2020magnetic}
V.~M. Kuchkin, B.~Barton-Singer, F.~N. Rybakov, S.~Bl\"ugel, B.~J. Schroers and
  N.~S. Kiselev, \emph{{Magnetic skyrmions, chiral kinks and holomorphic
  functions}}, \href{https://doi.org/10.1103/PhysRevB.102.144422}{\emph{Phys.
  Rev. B} {\bfseries 102} (2020) 144422}
  [\href{https://arxiv.org/abs/2007.06260}{{\ttfamily 2007.06260}}].

\bibitem{ross2021skyrmion}
C.~Ross, N.~Sakai and M.~Nitta, \emph{{Skyrmion interactions and lattices in
  chiral magnets: analytical results}},
  \href{https://doi.org/10.1007/JHEP02(2021)095}{\emph{JHEP} {\bfseries 02}
  (2021) 095} [\href{https://arxiv.org/abs/2003.07147}{{\ttfamily
  2003.07147}}].

\bibitem{hill2021chiral}
D.~Hill, V.~Slastikov and O.~Tchernyshyov, \emph{{Chiral magnetism: a geometric
  perspective}},
  \href{https://doi.org/10.21468/SciPostPhys.10.3.078}{\emph{SciPost Phys.}
  {\bfseries 10} (2021) 078}.

\bibitem{schroers2021solvable}
B.~Schroers, \emph{{Solvable Models of Magnetic Skyrmions}},  in \emph{{11th
  International Symposium on Quantum Theory and Symmetries}}, 10, 2019,
  \href{https://arxiv.org/abs/1910.13907}{{\ttfamily 1910.13907}},
  \href{https://doi.org/10.1007/978-3-030-55777-5_50}{DOI}.

\bibitem{Amari:2022boe}
Y.~Amari, Y.~Akagi, S.~B. Gudnason, M.~Nitta and Y.~Shnir, \emph{{CP2 skyrmion
  crystals in an SU(3) magnet with a generalized Dzyaloshinskii-Moriya
  interaction}},
  \href{https://doi.org/10.1103/PhysRevB.106.L100406}{\emph{Phys. Rev. B}
  {\bfseries 106} (2022) L100406}
  [\href{https://arxiv.org/abs/2204.01476}{{\ttfamily 2204.01476}}].

\bibitem{Hanada:2023lnm}
F.~Hanada and N.~Sawado, \emph{{A baby Skyrme model with anisotropic DM
  interaction: Compact skyrmions revisited}},
  \href{https://doi.org/10.1016/j.nuclphysb.2023.116377}{\emph{Nucl. Phys. B}
  {\bfseries 996} (2023) 116377}
  [\href{https://arxiv.org/abs/2303.15751}{{\ttfamily 2303.15751}}].

\bibitem{fert2013skyrmions}
A.~Fert, V.~Cros and J.~Sampaio, \emph{Skyrmions on the track},
  \href{https://doi.org/10.1038/nnano.2013.29}{\emph{Nature Nanotechnology}
  {\bfseries 8} (2013) 152}.

\bibitem{fert2017magnetic}
A.~Fert, N.~Reyren and V.~Cros, \emph{Magnetic skyrmions: advances in physics
  and potential applications},
  \href{https://doi.org/10.1038/natrevmats.2017.31}{\emph{Nature Reviews
  Materials} {\bfseries 2} (2017) 17031}.

\bibitem{muhlbauer2009skyrmion}
S.~M\"uhlbauer, B.~Binz, F.~Jonietz, C.~Pfleiderer, A.~Rosch, A.~Neubauer
  et~al., \emph{{Skyrmion Lattice in a Chiral Magnet}},
  \href{https://doi.org/10.1126/science.1166767}{\emph{Science} {\bfseries 323}
  (2009) 1166767}.

\bibitem{Munzer2010}
W.~M\"unzer, A.~Neubauer, T.~Adams, S.~M\"uhlbauer, C.~Franz, F.~Jonietz
  et~al., \emph{Skyrmion lattice in the doped semiconductor
  ${\text{fe}}_{1\ensuremath{-}x}{\text{co}}_{x}\text{Si}$},
  \href{https://doi.org/10.1103/PhysRevB.81.041203}{\emph{Phys. Rev. B}
  {\bfseries 81} (2010) 041203}.

\bibitem{Yu2010}
X.~Z. Yu, Y.~Onose, N.~Kanazawa, J.~H. Park, J.~H. Han, Y.~Matsui et~al.,
  \emph{Real-space observation of a two-dimensional skyrmion crystal},
  \href{https://doi.org/10.1038/nature09124}{\emph{Nature} {\bfseries 465}
  (2010) 901}.

\bibitem{Seki2012}
S.~Seki, X.~Z. Yu, S.~Ishiwata and Y.~Tokura, \emph{Observation of skyrmions in
  a multiferroic material},
  \href{https://doi.org/10.1126/science.1214143}{\emph{Science} {\bfseries 336}
  (2012) 198}.

\bibitem{Yu2011}
X.~Z. Yu, N.~Kanazawa, Y.~Onose, K.~Kimoto, W.~Z. Zhang, S.~Ishiwata et~al.,
  \emph{Near room-temperature formation of a skyrmion crystal in thin-films of
  the helimagnet fege}, \href{https://doi.org/10.1038/nmat2916}{\emph{Nature
  Materials} {\bfseries 10} (2011) 106}.

\bibitem{nagaosa2013topological}
N.~Nagaosa and Y.~Tokura, \emph{Topological properties and dynamics of magnetic
  skyrmions}, \href{https://doi.org/10.1038/nnano.2013.243}{\emph{Nature
  Nanotechnology} {\bfseries 8} (2013) 899}.

\bibitem{Tokura2021}
Y.~Tokura and N.~Kanazawa, \emph{Magnetic skyrmion materials},
  \href{https://doi.org/10.1021/acs.chemrev.0c00297}{\emph{Chemical Reviews}
  {\bfseries 121} (2021) 2857}.

\bibitem{everschor2018perspective}
K.~Everschor-Sitte, J.~Masell, R.~M. Reeve and M.~Kläui, \emph{{Perspective:
  Magnetic skyrmions—Overview of recent progress in an active research
  field}}, \href{https://doi.org/10.1063/1.5048972}{\emph{Journal of Applied
  Physics} {\bfseries 124} (2018) 240901}.

\bibitem{luo2021skyrmion}
S.~Luo and L.~You, \emph{{Skyrmion devices for memory and logic applications}},
  \href{https://doi.org/10.1063/5.0042917}{\emph{APL Materials} {\bfseries 9}
  (2021) 050901}.

\bibitem{psaroudaki2023skyrmion}
C.~Psaroudaki, E.~Peraticos and C.~Panagopoulos, \emph{{Skyrmion qubits:
  Challenges for future quantum computing applications}},
  \href{https://doi.org/10.1063/5.0177864}{\emph{Applied Physics Letters}
  {\bfseries 123} (2023) 260501}.

\bibitem{dzyaloshinsky1958thermodynamic}
I.~Dzyaloshinsky, \emph{A thermodynamic theory of “weak” ferromagnetism of
  antiferromagnetics},
  \href{https://doi.org/https://doi.org/10.1016/0022-3697(58)90076-3}{\emph{Journal
  of Physics and Chemistry of Solids} {\bfseries 4} (1958) 241}.

\bibitem{moriya1960anisotropic}
T.~Moriya, \emph{Anisotropic superexchange interaction and weak
  ferromagnetism}, \href{https://doi.org/10.1103/PhysRev.120.91}{\emph{Phys.
  Rev.} {\bfseries 120} (1960) 91}.

\bibitem{moriya1960new}
T.~Moriya, \emph{New mechanism of anisotropic superexchange interaction},
  \href{https://doi.org/10.1103/PhysRevLett.4.228}{\emph{Phys. Rev. Lett.}
  {\bfseries 4} (1960) 228}.

\bibitem{Bak_1980}
P.~Bak and M.~H. Jensen, \emph{Theory of helical magnetic structures and phase
  transitions in mnsi and fege},
  \href{https://doi.org/10.1088/0022-3719/13/31/002}{\emph{Journal of Physics
  C: Solid State Physics} {\bfseries 13} (1980) L881}.

\bibitem{Bolognesi:2014ova}
S.~Bolognesi and W.~Zakrzewski, \emph{{Baby Skyrme Model, Near-BPS
  Approximations and Supersymmetric Extensions}},
  \href{https://doi.org/10.1103/PhysRevD.91.045034}{\emph{Phys. Rev. D}
  {\bfseries 91} (2015) 045034}
  [\href{https://arxiv.org/abs/1407.3140}{{\ttfamily 1407.3140}}].

\bibitem{Gisiger:1996vb}
T.~Gisiger and M.~B. Paranjape, \emph{{Solitons in a baby Skyrme model with
  invariance under volume / area preserving diffeomorphisms}},
  \href{https://doi.org/10.1103/PhysRevD.55.7731}{\emph{Phys. Rev. D}
  {\bfseries 55} (1997) 7731}
  [\href{https://arxiv.org/abs/hep-ph/9606328}{{\ttfamily hep-ph/9606328}}].

\bibitem{Adam:2009px}
C.~Adam, P.~Klimas, J.~Sanchez-Guillen and A.~Wereszczynski, \emph{{Compact
  baby skyrmions}},
  \href{https://doi.org/10.1103/PhysRevD.80.105013}{\emph{Phys. Rev. D}
  {\bfseries 80} (2009) 105013}
  [\href{https://arxiv.org/abs/0909.2505}{{\ttfamily 0909.2505}}].

\bibitem{Adam:2010jr}
C.~Adam, T.~Romanczukiewicz, J.~Sanchez-Guillen and A.~Wereszczynski,
  \emph{{Investigation of restricted baby Skyrme models}},
  \href{https://doi.org/10.1103/PhysRevD.81.085007}{\emph{Phys. Rev. D}
  {\bfseries 81} (2010) 085007}
  [\href{https://arxiv.org/abs/1002.0851}{{\ttfamily 1002.0851}}].

\bibitem{Adam:2010fg}
C.~Adam, J.~Sanchez-Guillen and A.~Wereszczynski, \emph{{A Skyrme-type proposal
  for baryonic matter}},
  \href{https://doi.org/10.1016/j.physletb.2010.06.025}{\emph{Phys. Lett. B}
  {\bfseries 691} (2010) 105}
  [\href{https://arxiv.org/abs/1001.4544}{{\ttfamily 1001.4544}}].

\bibitem{Klimas:2023zxm}
P.~Klimas, L.~C. Kubaski, N.~Sawado and S.~Yanai, \emph{{Gauged compact Q-balls
  and Q-shells in a multi-component $CP^N$ model}},
  \href{https://arxiv.org/abs/2311.13076}{{\ttfamily 2311.13076}}.

\bibitem{Klimas:2017eft}
P.~Klimas and L.~R. Livramento, \emph{{Compact Q-balls and Q-shells in CPN type
  models}}, \href{https://doi.org/10.1103/PhysRevD.96.016001}{\emph{Phys. Rev.
  D} {\bfseries 96} (2017) 016001}
  [\href{https://arxiv.org/abs/1704.01132}{{\ttfamily 1704.01132}}].

\bibitem{Adam:2008rf}
C.~Adam, P.~Klimas, J.~Sanchez-Guillen and A.~Wereszczynski, \emph{{Compact
  gauge K vortices}},
  \href{https://doi.org/10.1088/1751-8113/42/13/135401}{\emph{J. Phys. A}
  {\bfseries 42} (2009) 135401}
  [\href{https://arxiv.org/abs/0811.4503}{{\ttfamily 0811.4503}}].

\bibitem{Doring2017}
L.~D{\"o}ring and C.~Melcher, \emph{Compactness results for static and dynamic
  chiral skyrmions near the conformal limit},
  \href{https://doi.org/10.1007/s00526-017-1172-2}{\emph{Calculus of Variations
  and Partial Differential Equations} {\bfseries 56} (2017) 60}.

\bibitem{walton2020geometry}
E.~Walton, \emph{{On the geometry of magnetic Skyrmions on thin films}},
  \href{https://doi.org/10.1016/j.geomphys.2020.103802}{\emph{J. Geom. Phys.}
  {\bfseries 156} (2020) 103802}
  [\href{https://arxiv.org/abs/1908.08428}{{\ttfamily 1908.08428}}].

\bibitem{10.21468/SciPostPhys.8.6.086}
S.~Komineas and N.~Papanicolaou, \emph{{Traveling skyrmions in chiral
  antiferromagnets}},
  \href{https://doi.org/10.21468/SciPostPhys.8.6.086}{\emph{SciPost Phys.}
  {\bfseries 8} (2020) 086}.

\bibitem{Leese:1989gi}
R.~A. Leese, M.~Peyrard and W.~J. Zakrzewski, \emph{{Soliton Scatterings in
  Some Relativistic Models in (2+1)-dimensions}},
  \href{https://doi.org/10.1088/0951-7715/3/3/011}{\emph{Nonlinearity}
  {\bfseries 3} (1990) 773}.

\bibitem{Zheng2021}
F.~Zheng, F.~N. Rybakov, N.~S. Kiselev, D.~Song, A.~Kov{\'a}cs, H.~Du et~al.,
  \emph{Magnetic skyrmion braids},
  \href{https://doi.org/10.1038/s41467-021-25389-7}{\emph{Nature
  Communications} {\bfseries 12} (2021) 5316}.

\end{thebibliography}\endgroup
\end{document}